 \documentclass[preprint,12pt,authoryear]{elsarticle}

\usepackage{amssymb}
\usepackage{amsmath}
\usepackage{booktabs}
\usepackage{tikz}
\usepackage{hyperref}
\usepackage{afterpage}
\usepackage{bbding}
\usepackage[utf8]{inputenc}
\usepackage[T1]{fontenc}
\usepackage{tgtermes}
\usepackage{float}

\newcommand{\greendot}{\protect\greendotaux}
\newcommand{\greendotaux}{
\begin{tikzpicture}
\filldraw[fill=green,draw=green] circle (2pt);
\end{tikzpicture}
}

\newcommand{\greencircle}{\protect\greencircleaux}
\newcommand{\greencircleaux}{
\begin{tikzpicture}
\filldraw[fill=white,draw=green] circle (2pt);
\end{tikzpicture}
}

\newcommand{\yellowdot}{\protect\yellowdotaux}
\newcommand{\yellowdotaux}{
\begin{tikzpicture}
\filldraw[fill=yellow,draw=yellow] circle (2pt);
\end{tikzpicture}
}

\newcommand{\purpledot}{\protect\purpledotaux}
\newcommand{\purpledotaux}{
\begin{tikzpicture}
\filldraw[fill=purple,draw=purple] circle (2pt);
\end{tikzpicture}
}

\newcommand{\orangedot}{\protect\orangedotaux}
\newcommand{\orangedotaux}{
\begin{tikzpicture}
\filldraw[fill=orange,draw=orange] circle (2pt);
\end{tikzpicture}
}

\newcommand{\bluedot}{\protect\bluedotaux}
\newcommand{\bluedotaux}{
\begin{tikzpicture}
\filldraw[fill=blue,draw=blue] circle (2pt);
\end{tikzpicture}
}

\newcommand{\reddot}{\protect\reddotaux}
\newcommand{\reddotaux}{
\begin{tikzpicture}
\filldraw[fill=red,draw=red] circle (2pt);
\end{tikzpicture}
}

\journal{Medical Image Analysis}

\begin{document}

\begin{frontmatter}



\title{Patherea: Cell Detection and Classification for the 2020s} 


\address[ul_fri_address]{Faculty of Computer and Information Science, University of Ljubljana, Večna pot 113, 1000 Ljubljana, Slovenia}
\address[ip_address]{Institute of Pathology, Faculty of Medicine, University of Ljubljana, Korytkova 2, 1000 Ljubljana, Slovenia}
\address[oi_address]{Institute of Oncology Ljubljana, Zaloška cesta 2, 1000 Ljubljana, Slovenia}

\author[ul_fri_address]{Dejan Štepec}
\cortext[cor1]{Corresponding author. Email: dejan.stepec@patherea.ai}
\author[ip_address]{Maja Jerše}
\author[oi_address]{Snežana Đokić}
\author[ip_address]{Jera Jeruc}
\author[ip_address]{Nina Zidar}
\author[ul_fri_address]{Danijel Skočaj}


\begin{abstract}
We present Patherea, a unified framework for point-based cell detection and classification that enables the development and fair evaluation of state-of-the-art methods. To support this, we introduce a large-scale dataset that replicates the clinical workflow for Ki-67 proliferation index estimation. Our method directly predicts cell locations and classes without relying on intermediate representations. It incorporates a hybrid Hungarian matching strategy for accurate point assignment and supports flexible backbones and training regimes, including recent pathology foundation models. Patherea achieves state-of-the-art performance on public datasets—Lizard, BRCA-M2C, and BCData—while highlighting performance saturation on these benchmarks. In contrast, our newly proposed Patherea dataset presents a significantly more challenging benchmark. Additionally, we identify and correct common errors in current evaluation protocols and provide an updated benchmarking utility for standardized assessment. The Patherea dataset and code are publicly available to facilitate further research and fair comparisons.

\end{abstract}




\end{frontmatter}



\section{Introduction}
\label{sec1}

The emergence of whole-slide scanners has enabled the scanning of a complete microscopic tissue slide and the creation of a single high-resolution digital file - a gigapixel whole slide image (WSI). The so-called digital pathology represents recent digitalization efforts in the medical field of pathology to omit the need for traditional diagnostics to be performed under the microscope~\citep{mukhopadhyay2018whol}. However, with the emergence of whole slide scanners and progress in AI and computer vision, it is simultaneously undergoing another big transformation, shifting from digital to computational pathology – the analysis of digitized data using AI~\citep{bera2019artificial,cui2021artificial}. Various computer vision approaches have been applied for histopathology analysis and diagnostics with proven comparative performance with the pathologists~\citep{bejnordi2017diagnostic,bulten2022artificial}.

Cancer diagnostics is traditionally performed by pathologists under the microscope, which includes quantifying the immunohistochemical expression of various proteins on specially stained tissue samples. In practice, this means counting hundreds or thousands of cells of a particular class and interest (e.g., positively, and negatively stained tumor cells) which forms a basis to derive prognostic scoring markers~\citep{dowsett2011assessment}. Ki-67 proliferation index represents one of the most widely used prognostic markers in different cancer types (e.g., breast cancer, neuroendocrine tumors, lymphomas, sarcomas) and identifies the proportion of tumor cells in the proliferation phase (i.e., tumor growth rate). Computing such a scoring is a highly time-consuming manual work for a pathologist, however it enables classification and grading of tumors, evaluation of their malignant potential, and is the basis for determining an effective treatment. In practice, less accurate approaches are used~\cite{mikami2013interobserver,polley2013international}, which leads to lower interobserver concordance and reproducibility.

Automated cell detection and classification are essential for efficient and reproducible diagnostics, and they form a fundamental topic in histopathology image analysis~\cite{xing2016robust}. Supervised learning approaches depend on expertly labeled datasets, which are expensive and time-consuming to obtain at scale compared to natural image datasets in computer vision. Most methods rely on point-based annotations, which are more accessible than bounding boxes or segmentation masks and typically sufficient for diagnostic applications. Recent deep learning approaches~\citep{xie2015beyond,xie2018efficient,lee2021differential,abousamra2021multi,huang2023affine,huang2023prompt,pina2024cell} employ intermediate representations to regress and classify cell centers or adapt complex DETR-based~\citep{carion2020end} object detection architectures for point annotations. In contrast, our approach provides direct point-to-point mapping using standard computer vision backbones~\citep{dosovitskiy2021an,liu2022convnet}, offering a simpler architecture with faster convergence and compatibility with existing large-scale pathology foundation models.


Generating labeled data for different possible diseases is not only prohibitively expensive and time-consuming, but in the medical domain, often impossible due to low occurrence and natural and technical variability present in the samples. On the other hand, diagnostics is performed daily around the world, with workflows being digitized and unlabeled samples being stored in large quantities. Similar to general language and vision domains~\citep{radford2019language,radford2021learning,he2022masked,oquab2024dinov}, pathology foundation models have been recently presented~\citep{filiot2023scaling,chen2024towards,vorontsov2023virchow,dippel2024rudolfv,xu2024whole,nechaev2024hibou} that utilize vast amounts of unlabeled data from up to millions of samples from various diseases, stainings and instruments.

Limited labeled training data is essential both for fine-tuning models on specific downstream tasks and for benchmarking different approaches. However, publicly available datasets for cell detection and classification in pathology are scarce. Most existing datasets are based on standard hematoxylin and eosin (H\&E) staining and contain, at most, a few tens of thousands of annotated cell centers. These annotations are often produced semi-automatically and limited to small image patches, which does not reflect the real-world workflow of pathologists~\citep{sirinukunwattana2016locality,huang2020bcdata,graham2021lizard,abousamra2021multi,ryu2023ocelot}. Furthermore, the use of inconsistent evaluation metrics and implementation details across studies makes it challenging to compare the performance of different methods in a meaningful way.


To summarize, we make the following contributions:

\begin{itemize}
  \item We propose a novel end-to-end point-based cell detection and classification architecture based on Vision Transformers that completely removes the need for pre-processing and post-processing, enabling fast convergence and utilization of domain-specific foundation models.
  \item We identify and address the erroneous calculation of performance metrics in point-based object detection in the domain of pathology and provide a benchmarking framework that addresses the identified issues and enables a fair comparison of different approaches.
  \item We present the largest manually labeled point-based Ki-67 dataset, annotated by four senior pathologists on full-resolution WSIs following clinical protocols. Our dataset reveals performance saturation on existing public datasets and poses significantly greater challenges for current state-of-the-art approaches.
  \item We demonstrate significant improvements over state-of-the-art methods on both existing and newly proposed cell detection and classification datasets, achieving superior F1 scores while maintaining faster training and inference speeds.
  \item We make our Patherea framework (dataset, code) publicly available for research use\footnote{\href{https://github.com/patherea/patherea-p2p}{https://github.com/patherea/patherea-p2p}}.
\end{itemize}

\section{Related Work}
\label{sec2}

\subsection{Point-based Cell Detection}
\label{sec2_foundation}

Cell detection and classification in histopathology is a challenging problem due to variance in cell shape and appearance. This is further exacerbated by overlapping cells, staining artifacts, scanning artifacts, out-of-focus regions, and in general, the sheer amount of cells present in a selected field-of-view. This represents a significant challenge even for trained pathologists when performing diagnostics under the miscroscope~\citep{mikami2013interobserver,polley2013international}. While limited success can be achieved with conventional image processing approaches, that are included in some of the open-source pathology software's~\cite{bankhead2017qupath}, we limit ourselves to learning-based approaches.

The early approaches were based on classifying small patches centered on the ground-truth cell-center locations of whether the center of the patch belongs to the foreground or background, or by learning a distance function for each pixel in a centered patch~\citep{kainz2015you,sirinukunwattana2016locality}. The intermediate representation in the form of different distance-based functions has later evolved by taking into account the neighboring context and a direct regression on full-sized patches~\citep{xie2018efficient,huang2020bcdata}. MCSpatNet~\citep{abousamra2021multi} has recently introduced Ripley's K-function that represents an expected number of neighbors of a specific type, around the limited vicinity of the target cell location. By learning to predict the vectorized form of a K-function, the model essentially learns the spatial representation. These approaches regress to an intermediate representation, which needs to be pre-computed for training the regression model. In inference, the predicted intermediate representation needs to be post-processed (e.g., with non-maxima suppression) to get the individual cell locations and cell types. Such approaches are better-suited for cell-counting, while error-prone in cell localization tasks.

Cell detection and classification represents a similar task to crowd counting and localization~\citep{li2018csrnet,song2021rethinking} in the broader computer vision domain. P2PNet~\citep{song2021rethinking} represents a purely point-based approach, without the need for intermediate representations. This is achieved by assigning an optimal target to each proposal candidate in a one-to-one fashion using the Hungarian algorithm. A similar approach is utilized in Detection-Transformers (DETRs), which were recently proposed for bounding-box-based object detection~\citep{carion2020end}. DETR-based approaches were recently adapted for point-based detection and applied to pathology~\citep{huang2023affine,huang2023prompt}. From the implementation perspective, the standard bounding-box DETR is utilized with $1\times 1\;px$ bounding box centered on ground-truth point-based annotations. The naive implementation diminishes the predictive power of visual features near the cell center, prompting the development of various advancements to address this limitation. ACFormer~\citep{huang2023affine} utilizes a DETR-based architecture and proposes an Affine-Consistent-Transformer by using local and global networks to enhance the spatial scale consistency. PGT~\citep{huang2023prompt} is also based on DETR and introduces a learnable Grouping Transfer that leverages the similarity between nuclei and their cluster representation to take into account the spatial context, similar to MCSpatNet~\citep{abousamra2021multi}.

Point-based cell detection can be viewed as a simplified alternative to instance segmentation methods, which have become prevalent in the pathology domain~\citep{graham2019hover,graham2023one,cellvit,cellvit++}. While instance segmentation provides detailed cellular boundaries, it relies on labor-intensive pixel-level annotations that are often impractical to obtain at scale. In contrast, point annotations are significantly cheaper and faster to acquire. Moreover, for many downstream pathology tasks--such as cell counting, phenotyping, or spatial analysis—the exact cell shape offers limited additional utility beyond the cell’s location. As such, point-based methods offer a more annotation-efficient solution while still meeting the requirements of many pathology applications.

Our proposed Patherea-P2P approach (Section~\ref{sec3}) builds upon P2PNet~\citep{song2021rethinking} and introduces a Hybrid Hungarian Matching to increase the amount of positive supervision compared to the vanilla one-to-one matching utilized in P2PNet and DETR-based approaches. Implicitly, this also incorporates contextual information. We also design our architecture around a standard visual backbone with a lightweight head, in comparison with object detection specific DETR-based architectures, which limit the use of large pre-trained foundation models.

\subsection{Datasets}
\label{sec2_datasets}

Supervised deep-learning approaches require lots of labeled training data. Acquiring labeled data in the medical imaging domain is particularly challenging, as the samples need to be labeled by domain experts, which represents the main difference in comparison with natural images. Providing fine-grained annotations (e.g., point annotations, bounding boxes, segmentation masks) represents an even greater challenge due to the time-consuming annotation process. Larger datasets in digital pathology are thus mostly focused on patch-based classification~\citep{litjens20181399,sirinukunwattana2017gland,kather2018100,bulten2022artificial}, where particular (larger) tissue regions are delineated with polygons in WSIs, with patches being extracted and labeled based on the annotated source polygon region.

Cell (nuclei) detection and classification requires much finer-grained labels. A selection of the most prominent datasets is presented in Table~\ref{tbl:datasets_related}.

\setlength{\tabcolsep}{1\tabcolsep}
\begin{table}[ht!]
\centering
\caption{Characteristics of different cell detection and classification datasets. Cell segmentation datasets (*) can also be utilized. Datasets are ordered by their publication date. The WSI column indicates whether the dataset was acquired using a digital slide scanner. The manual column specifies whether the dataset was annotated fully manually.}
\begin{tabular}{@{}llllllll@{}}
\toprule
            & Stain & Organs & Magn.  & Cells & Classes & WSI & Manual     \\ \toprule
BM     & H\&E  & 1     & 20x & 4,205 & 1 & \Checkmark & \Checkmark    \\
CRC    & H\&E  & 1     & 20x & 22,444 & 4 & \Checkmark & \Checkmark    \\
CoNSeP*       & H\&E  & 1 & 40x     & 24,319 & 4 & \Checkmark & \Checkmark    \\
PanNuke*       & H\&E  & 19     & 20/40x & 189,744 & 5 & \Checkmark & \XSolid    \\
BCData       & Ki-67 & 1     & 40x & 181,074 & 2 & \XSolid & \Checkmark    \\
Lizard*       & H\&E  & 1     & 20x & 495,179 & 6 & \Checkmark & \XSolid    \\
BRCA-M2C       & H\&E  & 1     & 20x & 30,638 & 3 & \Checkmark & \Checkmark    \\
OCELOT       & H\&E  & 6     & 40x & 114,700 & 2 & \Checkmark & \Checkmark    \\ \toprule
Patherea (\textbf{ours})       & Ki-67  & 7     & 40x & 202,807 & 4 & \Checkmark & \Checkmark    \\ \bottomrule
\end{tabular}
\label{tbl:datasets_related}
\end{table}

First point-based datasets (e.g., BM~\citep{kainz2015you}, CRC~\citep{sirinukunwattana2016locality}) were manually labeled single-organ datasets with a limited number of labeled cell nuclei and cell types. Some of the datasets are labeled at 20x objective magnification, which significantly reduces the level of morphology present in cellular structures, which can hinder effective cell detection and classification. Access to large-scale 40x objective magnification samples is limited, as hospitals mostly archive 20x samples, due to storage constraints. Cell segmentation and detection datasets were also introduced (e.g., CoNSeP~\citep{graham2019hover}, PanNuke~\citep{gamper2019pannuke}, Lizard~\citep{graham2021lizard}) which usually also have point-based annotations, or can be manually computed as centroids of the segmentation masks.

Cell segmentation datasets~\citep{gamper2019pannuke,graham2021lizard} were mostly labeled in a semi-automatic fashion, by manually labeling a small fraction of the samples, training the model, applying the model on new samples, followed by a manual step of (frequently limited) expert-level correction. The Lizard dataset~\citep{graham2021lizard} represents the largest dataset with close to 500k labeled nuclei. A slightly extended version of the Lizard dataset was also part of the recent CoNIC nuclear detection, segmentation, classification and counting challenge~\citep{graham2024conic}.

BCData~\citep{huang2020bcdata}, BRCA-M2C~\citep{abousamra2021multi} and \\OCELOT~\citep{ryu2023ocelot} are more recent datasets that were specifically designed for the task of point-based cell detection and classification tasks. BCData~\citep{huang2020bcdata} represents a similar task to our proposed Patherea-Breast dataset (Section~\ref{sec4}), and the only dataset that uses stains beyond the standard H\&E, but was collected using a standard microscope, with an attached camera. OCELOT additionally provides a larger field-of-view pair of cell and tissue annotations~\citep{ryu2023ocelot}.

Our proposed Patherea dataset to the best of our knowledge represents the largest, fully-manually-labeled point-based cell detection and classification dataset acquired at 40x objective magnification. In comparison with related work (Table~\ref{tbl:datasets_related}), we acquire the dataset in a way that mimics the traditional clinical workflow with a microscope, with expert pathologists interactively using the full WSIs at different magnifications to label the regions and cellular structures of interest.
 
\subsection{Pathology Foundation Models}
\label{sec2_foundation}

Transfer learning from a supervised pre-trained ImageNet~\citep{deng2009imagenet,yosinski2014transferable} backbone has been a widely established approach in general computer vision. Recently, transfer-learning from a self-supervised pre-training showed promising results in various downstream tasks~\citep{ericsson2021well,goldblum2024battle}. This is even more evident in the medical imaging domain~\citep{azizi2021big}, where large supervised pre-trained models are not available for a specific domain. The benefits of using self-supervised pre-training for fine-grained downstream tasks (e.g., detection, segmentation) have been less evident~\citep{goldblum2024battle}, especially with earlier, contrastive-based approaches~\citep{chen2020simple,caron2021emerging}. Lately, masked-image-modeling has been proposed~\citep{xie2022simmim,he2022masked,oquab2024dinov,tian2023designing}, which pre-text task is much more suitable for fine-grained tasks.

Similar to the general computer vision domain, pathology foundation models have been recently proposed~\citep{filiot2023scaling,chen2024towards,vorontsov2023virchow,dippel2024rudolfv,xu2024whole,nechaev2024hibou} that utilize vast amounts of digitized pathology data from various organs, stains and instruments. Most of the models utilize the standard H\&E data, which is available in abundance. Recent foundation models~\citep{dippel2024rudolfv,nechaev2024hibou} have also incorporated different stains. Most of the models are trained on proprietary training data and are not publicly available. They are predominantly evaluated on various downstream clinical tasks, where a significant boost is observed in comparison with ImageNet pre-trained models~\citep{campanella2024clinical}. Task-specific approaches that utilize foundation models have also started to appear---such as recently presented CellViT$^{++}$~\citep{cellvit++} instance segmentation-based approach that reduces the need for segmentation masks in training.

We utilize a recently provided open-source Hibou pathology foundation model\-~\citep{nechaev2024hibou}, based on the ViT-B backbone and apply it to the task of point-based cell detection and classification. This presents a novel application of an existing foundation model to the specific downstream task of point-based cell detection and classification.

\section{Methods}
\label{sec3}

\begin{figure}[t]
\centering
\includegraphics[width=\textwidth,height=\textheight,keepaspectratio]{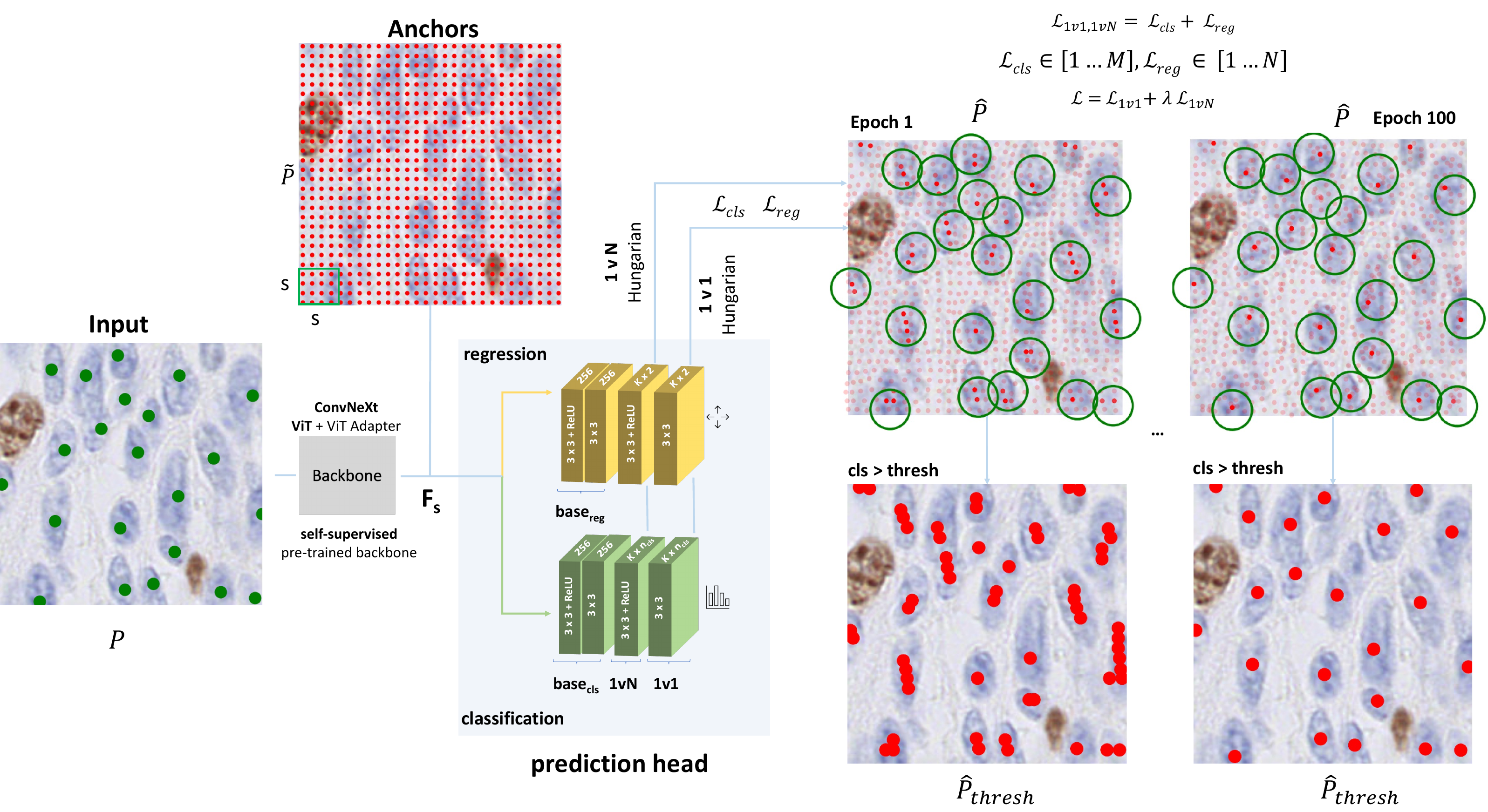}
\caption{The proposed Patherea-P2P architecture. The green dots ($\greendot$) and circles ($\greencircle$) represent the ground-truth annotations for one particular cell class. Anchors ($\reddot$) are initialized in a regular grid and get moved around using the regression loss (yellow head) and classified using the classification head (green). The alpha channel in $\mathcal{\hat{P}}$ reflects anchor confidence, clipped at 0.2 to keep low-confidence anchors visible.}
\label{fig1}
\end{figure}

The proposed Patherea-P2P architecture is presented in Figure~\ref{fig1}. The overall architecture was inspired by the crowd counting and localization Point-to-Point Network (P2PNet)~\citep{song2021rethinking}, which was further extended with a novel Hybrid Hungarian-based regression loss, classification capability and modern building blocks that enable the use of self-supervised approaches to train or use existing foundation models tailored for digital pathology. The used notation mostly follows P2PNet~\citep{song2021rethinking}.

\subsection{P2P Network}
\label{subsec_p2p}

Patherea-P2P method training input represents a (histology tissue) patch with a set of $\mathcal{P} = \{p_i\}$ point-based annotations $p_i=(x_i,y_i,t_i), i \in \{1,...,N\}, t \in \{1,...,T\}$,  which represents a cell center location $(x_i,y_i)$ and a cell type $(t_i)$ for a particular cell $i$ - among $N$ labeled cells and  $T$ cell types in a given patch (Figure~\ref{fig1}, left, green dots for one particular cell class $t$). The trained model similarly predicts a set of points $\mathcal{\hat{P}} = \{\hat{p_j}\}$ in a given patch $\hat{p_j}=(\hat{x_j},\hat{y_j},c_j^{t\in T}), j \in \{1,...,M\}$, based on the $M$ proposal candidates (Figure~\ref{fig1}, top-right, red dots), with an additional confidence score $c_j^{t\in T}$ for each of the $T$ possible cell types, which we threshold to produce a final set of predictions (Figure~\ref{fig1}, bottom-right, red dots). Predicted points $\hat{p_j}$ should be as close as possible to ground-truth cell locations and of the same type $t_i$, while maximizing a confidence score $c_j$, for the task of cell localization. Similarly, the overall number of predicted number of cells ($\hat{N}=|c_j^{t\in T} > c_{thresh}^{t\in T}|$) should be as close as possible to the actual number of cells $N$ in a given patch, for the task of cell counting. Patherea-P2P combines and optimizes both tasks at the same time, while taking into account individual cell types.

The overall architecture consists of a backbone, which encodes an image patch to the embedding $\mathcal{F}_s$, where $s$ represents the downsampling stride (each positional embedding in $\mathcal{F}_s$ represents a $s \times s$ receptive field in the input patch). The lightweight regression head $\mathcal{H}_{reg}$ and classification head $\mathcal{H}_{cls}$ are attached to the backbone, which are executed in parallel. We start by a set of proposal candidate points $\mathcal{\tilde{P}}$ (Figure~\ref{fig1}, top, red dots), which are initialized in a regular grid along the input patch. This is achieved by initializing $K$ proposal candidates for each receptive field of size $s \times s$ in a $H_f \times W_f$ sized embedding $\mathcal{F}_s$ ($M=H_f*W_f*K$). Regression head $\mathcal{H}_{reg}$ then predicts the offsets $\Delta_{x,y}^{k\in K}$ for each proposal $k\in K$. Similarly, classification head $\mathcal{H}_{cls}$ predicts cell class for each of the $K$ proposal candidates.

For every ground truth target from $\mathcal{P}$, we need to assign a proposal from a set of predicted points $\mathcal{\hat{P}}$, using a one-to-one matching strategy. Point proposals are dynamically updated during training and there is no guarantee that the same point proposal will always be matched to the same ground truth proposal, especially in the early stages of training, where proposals compete with each other. But crucially, an optimal one-to-one matching must be found, where there is only one proposal candidate for each ground truth target. The matching process can be described with $\Omega(\mathcal{P},\mathcal{\hat{P}},\mathcal{D})$, where $\mathcal{D}$ represents a pair-wise matching cost matrix $N\times M$, that needs to be optimized to produce matchings that minimize the sum of individual matching pair-wise costs~(\ref{eq_hungarian_cost}):

\begin{equation}
\mathcal{D}(\mathcal{P},\mathcal{\hat{P}})=(\tau \|p_i - \hat{p_j}\|_2 - c_j^{t_i}),
\label{eq_hungarian_cost}
\end{equation}

\noindent where $\|\cdot\|_2$ denotes the $l_2$ distance between the matched ground truth cell location $p_i$ and matched predicted cell location $\hat{p_j}$. The predicted confidence score $c_j^{t_i}$ for the ground truth cell type $t_i$ is utilized to resolve potential conflicts where multiple proposals $p_j$ are of the same distance to the ground truth $p_i$. The confidence term also encourages the positive matches to have a higher confidence score. $\tau$ denotes a weight term to balance the pixel distance term. The Hungarian algorithm is used for $\Omega$ to solve the assignment problem.

Let $\xi=\Omega(\mathcal{P},\mathcal{\hat{P}},\mathcal{D})$ denote the optimal permutation $\{1,...,M\}$, such that $\mathcal{\hat{P}}_{pos}=\hat{p}_{\xi(i)}, i\in \{1,...,N\}$ represents the matched proposal candidate for ground truth $p_i$, while $\mathcal{\hat{P}}_{neg}=\hat{p}_{\xi(i)}, i\in \{N+1,...,M\}$ represents unmatched proposal candidates that are treated as background. The classification loss can thus be derived as follows ~(\ref{eq_cls_loss}):

\begin{equation}
\mathcal{L}_{cls}=-\frac{1}{M}\Biggl\{\sum_{i=1}^{M}\lambda_tt_i\log{c_{\xi(i)}^{t_i\in T}}\Biggr\},
\label{eq_cls_loss}
\end{equation}

\noindent where $\lambda_t$ denotes a cell type class weight for $T\in \{bg,t_1,...,t_{|T|}\}$ and $t_i$ denotes the ground truth cell type for the matching target $p_i$ and $|T|$ the number of all possible cell type classes. An additional background class $bg$ is introduced for $\mathcal{\hat{P}}_{neg}$.

Regression head $\mathcal{H}_{reg}$ is optimized using only matched targets $\mathcal{\hat{P}}_{pos}$ using the Euclidean loss:

\begin{equation}
\mathcal{L}_{reg}=\frac{1}{N}\sum_{i=1}^{N}\|p_i - \hat{p}_{\xi(i)}\|_2^2,
\label{eq_reg_loss}
\end{equation}

\noindent The combined loss is~(\ref{eq_combined_loss}):

\begin{equation}
\mathcal{L}_{1v1}=\mathcal{L}_{cls}+\lambda_{reg}\mathcal{L}_{reg},
\label{eq_combined_loss}
\end{equation}

\noindent where $\lambda_{reg}$ denotes the weight to balance the regression loss.

\subsection{Hybrid Hungarian Matching}
\label{subsec_hp2p}

With the P2P framework presented in Section~\ref{subsec_p2p}, only one proposal candidate is selected per ground truth target (one-to-one matching). Selected targets are supervised both, with the classification loss $\mathcal{L}_{cls}$~(\ref{eq_cls_loss}) and regression loss $\mathcal{L}_{reg}$~(\ref{eq_reg_loss}). Unmatched proposal candidates are only supervised with the classification loss. Given that $M > N$, often $M >> N$, there is a lack of positive supervision, especially for the regression loss. Additionally, in the earlier phases of the training, different proposal candidates are often selected and partially optimized, resulting in many candidates being localized and classified "correctly", however, in the later stages of training, these candidates are treated as negative examples. This is depicted in Figure~\ref{fig1} (top-right), where multiple proposal candidates are positioned on the cell (or even within the more restrictive ground-truth radius). The one-to-one matching described in~\ref{subsec_p2p} selects only the closest one by distance and highest confidence when in practice, multiple candidates can be equally good. This mixed supervision - "good" candidates are being treated as background - results in sub-optimal optimization and reduced performance.

We propose a hybrid matching scheme, that first performs one-to-many matching, followed by the standard one-to-one matching, as depicted in Figure~\ref{fig1}. The prediction head outputs are used to perform the one-to-many matching, while being further refined in the last layer with the standard one-to-one matching. This ensures more supervision signal from one-to-many optimization, while still enabling end-to-end point-based object detection, without the need for post-processing (e.g., non-maxima suppression).

We implement this by relaxing the Hungarian algorithm to enable multiple candidates to match the ground truth targets. This is achieved by replicating ground truth targets in a cost matrix $\mathcal{D}$ by a factor of $\beta$, where $\beta$ represents the number of candidate proposals per ground truth target. This allows the Hungarian algorithm to match multiple proposal candidates per ground truth target, with $\beta = 1$ representing the standard P2P framework presented in~\ref{subsec_p2p}. This results in a new cost matrix $\mathcal{\hat{D}}$ with dimensions ($N\times\beta)\times M$. The losses~(\ref{eq_cls_loss}) and~(\ref{eq_reg_loss}) are computed in the same manner, the only difference being more matched positive samples ($N=\beta\times N$), resulting in $\mathcal{L}_{1vN}$. The combined loss is then:

\begin{equation}
\mathcal{L}=\mathcal{L}_{1v1}+\lambda_{1vN}\mathcal{L}_{1vN},
\label{eq_combined_loss_hp2p}
\end{equation}

\noindent where $\lambda_{1vN}$ denotes the balancing weight between the one-to-one and one-to-many matchings.

\subsection{Foundation Model Support}
\label{subsec_backbone}

The proposed Patherea-P2P method removes the need for any intermediate representations on the input side, as well as any post-processing, thus making it fully end-to-end. Additionally, we design the architecture in a simplified manner, reducing the need for complex task-specific building blocks, beyond the light-weight heads. Most of the existing work~\citep{huang2023affine,huang2023prompt,pina2024cell} base their architecture on DETR-based frameworks. Vision, or even task-specific (e.g., object detection) approaches are usually more complex, but have usually achieved better performance due to the inherent inductive bias present in the architecture. We design the architecture around general-purpose architectures like ConvNext~\citep{liu2022convnet} and ViT~\citep{dosovitskiy2021an}, which directly support (multi-modal) pre-training in a self-supervised manner. We utilized ConvNext and ViT backbones, together with Feature Pyramid Networks (FPN)~\citep{lin2017feature}, which enables to use of higher-resolution features, beneficial for dense prediction tasks. For ViT, we additionally utilized ViT-Adapter~\citep{chen2023vision}, which allows plain ViT to achieve comparable performance to vision-specific transformers.

Masked image modelling can be used to train foundation models from scratch using MAE~\citep{he2022masked} for ViT or SparK~\citep{tian2023designing} for ConvNext. Alternatively, existing, open-source pathology foundation models~\citep{nechaev2024hibou,chen2024towards} can be used, which are most often only available for ViTs.

\section{Patherea Dataset}
\label{sec4}


We present, to the best of our knowledge, the largest public dataset for cell detection and classification on immunohistochemistry samples for the clinical problem of Ki-67 proliferation index estimation in neuroendocrine tumors (NETs) at different locations (small and large intestine, appendix, larynx, pharynx, lungs) and breast cancer. Both cancer types were selected due to well-standardized usage of Ki-67 as a classification/grading parameter by World Health Organization and other relevant standardization bodies and comparative studies~\citep{polley2013international,reid2015calculation,dowsett2011assessment,nielsen2021assessment}. There is also a significant lack of publicly available datasets with immunohistochemistry stainings, with vast majority of the public pathology datasets being the standard H\&E staining.



We collected 42 samples of NETs at the Institute of Pathology, Faculty of Medicine, University of Ljubljana and 29 samples of breast cancer at the Institute of Oncology Ljubljana. All the samples were scanned at x40 objective using Hamamatsu NanoZoomer S360. Ethical approvals were obtained from the National Medical Ethics Committee of the Republic of Slovenia for NETs (No. 0120-357/2023/23), as well as breast cancer (No. 0120-147/2024-2711-3).

To ensure clinical realism, we utilized whole slide images (WSIs) for annotation rather than pre-selected image patches, closely mimicking a digitized clinical workflow. The Digital Slide Archive (DSA) platform~\citep{gutman2017digital} was deployed in a cloud environment for managing and labeling WSIs, accessible to pathologists via a standard web browser.

Four pathologists participated in the annotation process. Each had unrestricted access to the entire WSI, allowing them to freely navigate, zoom across magnifications, and select regions of interest for annotation. This flexibility enabled the selection of more representative and diverse tissue areas, reducing sampling bias compared to fixed patch-based approaches. Annotated regions were outlined with polygons and all visible cells within these regions were exhaustively labeled with their respective cell types (Figure~\ref{fig:sup_platform}).

Four cell categories were defined: (i) positive tumor cells, (ii) negative tumor cells, (iii) positive non-tumor cells, and (iv) negative non-tumor cells. This labeling protocol reflects both diagnostic relevance and the clinical decision-making process (\ref{sec:supp_figures}).

We organized samples into three anatomically distinct groups: LNET (larynx, pharynx, and lung NET samples), GNET (gastrointestinal tract NET samples), and Breast (breast cancer samples). All annotations were performed by senior pathologists with an average of 25 years of clinical experience. In contrast to publicly available datasets that typically rely on junior pathologists or students with senior verification, our labeling approach ensures consistent annotation quality. Unlike large-scale datasets that employ semi-automatic labeling approaches~\citep{graham2021lizard}, where AI-proposed labels are subsequently verified, our dataset features entirely manual annotations, providing high-quality ground truth labels.

Different parts of the Patherea dataset, the number of samples and the number of different cells labeled are summarized in Table~\ref{tbl:patherea_dataset}. In total, more than 200k cells of different types were labelled across 100 samples, which, to the best of our knowledge represents the largest publicly available fully-manually labeled cell detection and classification dataset.

\setlength{\tabcolsep}{0.7\tabcolsep}
\begin{table}[ht!]
\centering
\caption{Statistics of the Patherea dataset. Different colors represent the colors used to represent different cell types across all the Figures in this manuscript. Breast-P1 and Breast-P2 represents the same dataset and ROIs that were labeled by two different pathologists.}
\begin{tabular}{@{}llllllll@{}}
\toprule
            & $N_{samples}$ & $N_{pos}^{\greendot}$ & $N_{neg}^{\yellowdot}$ & $N_{othr\_pos}^{\purpledot}$ & $N_{othr\_neg}^{\orangedot}$ & $\sum$ \\ \toprule
LNET        & 20 & 25,118 & 35,288 & 660 & 10,414 & 71,400  \\ \midrule
GNET        & 22 & 6,458  & 43,606 & 761 & 10,762  & 61,587  \\   \midrule
Breast-P1   & 29 & 9,665  & 17,187 & 908 & 8,913  & 36,673  \\
Breast-P2   & 29 & 8,813  & 18,583 &  274 & 5,477  & 33,147 \\ \midrule
$\sum$      & 100 & 50,054 & 114,664 & 2,603 & 35,486 & \textbf{202,807}  \\   \bottomrule
\end{tabular}
\label{tbl:patherea_dataset}
\end{table}

Nevertheless, there is some inherent bias with assigning cell types, including with senior pathologists. This is especially evident with borderline cases, such as other positive and negative cells. We have thus labeled the Breast part of the Patherea dataset by two pathologists. The same regions that were selected by the first pathologist were given to the second pathologist, which labeled all of the cells in the regions that were selected by the first pathologist. This is depicted in Table~\ref{tbl:patherea_dataset} by Breast-P1 and Breast-P2.

Annotated WSIs were then divided into $224\times 224$ patches and split into 3 folds for model training and evaluation in Section~\ref{sec6}. The dataset and fold splits are made publicly available\footnote{\href{https://huggingface.co/datasets/ds2268/patherea}{https://huggingface.co/datasets/ds2268/patherea}}.

\section{Evaluation Protocol}
\label{sec_5}

\begin{figure}[t]
\centering
\includegraphics[width=\textwidth,height=\textheight,keepaspectratio]{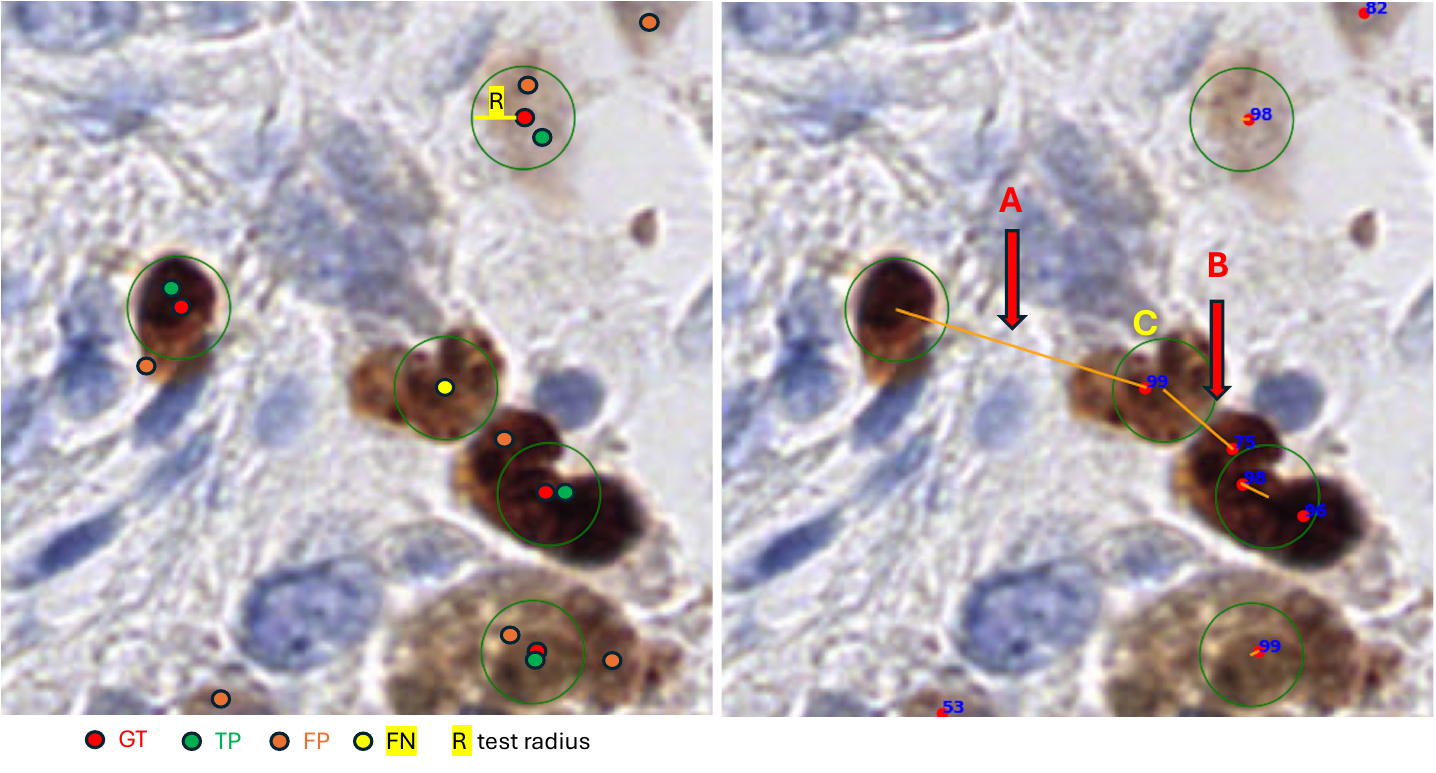}
\caption{Left: Visual representation of F1 metric computation based on the visual detections and circular ground truth regions. Right: Depiction of wrongly assigned matches when performing Hungarian one-to-one matching directly on the distance matrix. Numbers represent the predicted confidence.}
\label{fig3}
\end{figure}

As described in Section~\ref{sec4}, we only obtain point-based annotations and subsequently define a ground truth region as a circular region with radius $r$ centered at a cell center labeled by a pathologist. We can then match all the detected cell centroids with the corresponding pathologist annotations. We can then compute a per-class F1 score based on true-positives (TP), false-positives (FP) and false-negatives (FN), as depicted in Figure~\ref{fig3} - left and equation (\ref{eq_f1_eval}).

\begin{equation}
F1_{cls}=\frac{TP}{TP+\frac{1}{2}(FP+FN)}
\label{eq_f1_eval}
\end{equation}

The Hungarian algorithm is used by most of the established public datasets and benchmarks~\citep{xie2018efficient,graham2019hover,huang2023affine,huang2023prompt} to perform the one-to-one matching between the detections and labeled ground truth data. Some of the established benchmarks~\citep{graham2019hover} have performed the Hungarian matching incorrectly\footnote{\href{https://github.com/vqdang/hover\_net/blob/master/metrics/stats\_utils.py\#L393}{https://github.com/vqdang/hover\_net/blob/master/metrics/stats\_utils.py\#L393}}, which does not report the correct F1 scoring. Such incorrect scoring is then used by the following work~\citep{huang2023affine,huang2023prompt}, which prevents a fair comparison against different approaches and datasets, if not the same benchmarking code is used - which would still result in an underestimated F1 score. The problem arises due to the global cost optimization nature of the Hungarian algorithm, if a distance matrix is used directly to do the assignments. This is depicted in Figure~\ref{fig3} - right, which depicts the flawed nature of assignments.

Due to the global objective optimization, a distant detection, correctly detected for one of the ground truth cells (Figure~\ref{fig3}, C) is assigned to another cell that has no close detection (Figure~\ref{fig3}, A). A detection, close-by, but outside of the ground truth circular region is assigned to the first ground truth location (Figure~\ref{fig3}, B). After the Hungarian matching, filtering based on the ground truth radius $r$ is used. Due to the wrongly predicted point being matched to the first cell (Figure~\ref{fig3}, C), an additional false-negative is produced, despite the first cell being correctly detected in the first place.

This is the result of the global nature of the Hungarian optimization objective, which minimizes the total cost across all assignments (Equation~\ref{eq:hungarian_eval}):

\begin{equation}
\text{minimize: } \sum_{p_p \in P_p, p_g \in P_g} w(p_p, p_g) \cdot x(p_p, p_g)
\label{eq:hungarian_eval}
\end{equation}

\noindent where $w(p_p, p_g)$ is the weight (distance) between predicted point $p_p \in P_p$ and ground truth point $p_g \in P_g$, and $x(p_p, p_g)$ is a binary variable indicating assignment. A detection is correct if it falls within a specified radius $r$ of its corresponding ground truth. However, the Hungarian algorithm's global optimization can assign a detection to a distant ground truth.

The solution, correctly implemented in the broader domain of crowd counting and localization~\citep{wang2020nwpu}, involves applying the Hungarian algorithm to a Boolean matrix obtained by thresholding a distance matrix based on the ground truth radius $r$. Given two point sets of predicted points $P_p$ and ground truth points $P_g$, we first construct a weighted bipartite graph $G_{g,p}$, where edge weights represent the Euclidean distance between each predicted point $p_p \in P_p$ and each ground truth point $p_g \in P_g$. We then threshold this graph at radius $r$, retaining only edges with weights $\leq r$ to produce a Boolean match matrix. The Hungarian algorithm is applied to this thresholded graph to obtain the maximum bipartite matching, from which we compute true-positives (TP), false-positives (FP), and false-negatives (FN). We experimentally confirm the difference between the two approaches in~\ref{app2_hungarian}.



The alternative greedy approach of iterating across all the predicted points and ground truth data is used by some of the related work~\citep{abousamra2021multi}, which is also incorrect\footnote{\href{https://github.com/TopoXLab/MCSpatNet/blob/main/03\_eval\_localization\_fscore.py\#L17}{https://github.com/TopoXLab/MCSpatNet/blob/main/03\_eval\_localization\_fscore.py\#L17}}, by allowing multiple predictions within the ground truth radius count as true-positives (one-to-many matching), which overestimates the F1 score.


\section{Experiments}
\label{sec6}

In this section we report results on two established public datasets for point-based cell detection and classification - Lizard~\citep{graham2021lizard} and BRCA-M2C~\citep{abousamra2021multi} in Section~\ref{sec62}. We report results on our newly introduced Patherea dataset in Section~\ref{sec63}. We compare the proposed Patherea-P2P method introduced in Section~\ref{sec3} against the recent DETR-based approaches ACFormer~\citep{huang2023affine} and PGT~\citep{huang2023prompt}, as well as against more traditional approaches with an intermediate representation(s). For the traditional approaches, we compared against MCSpatNet~\citep{abousamra2021multi}, which additionally utilizes spatial context information. We also compared against our implementation of the approach from~\citep{xie2018efficient}, named FCRN++. This implementation was modernized with a ResNet-34 backbone and added support for cell classification, mostly following~\citep{lee2021differential}. We used the official code for ACFormer, PGT and MCSpatNet. When available, we also utilized their publicly available pre-trained weights for the Lizard (\ref{app1_lizard}) or BRCA-M2C (\ref{app2_brca}) datasets. Additionally, we also report results on BCData~\citep{huang2020bcdata} in~\ref{app2_bcdata}. F1 scores for Lizard, BRCA-M2C and Patherea datasets have been reported using the evaluation protocol described in Section~\ref{sec_5}.

\subsection{Implementation Details}
\label{sec61}

Patherea-P2P and FCRN++ methods were implemented in PyTorch, while the official code and parameters were used to reproduce MCSpatNet~\citep{abousamra2021multi}, ACFormer~\citep{huang2023affine} and PGT methods~\citep{huang2023prompt} on selected datasets. Patherea-P2P proposal candidates were initialized with $K=[2,2]$ and a Feature Pyramid Network (FPN)~\citep{lin2017feature} was used for ConvNext~\citep{liu2022convnet} and ViT~\citep{dosovitskiy2021an} backbones with number of features set to 256 for ConvNext-B and 768 for ViT-B. Only higher-resolution features at level 2 of the FPN were used as an input to $\mathcal{H}_{reg}$ and $\mathcal{H}_{cls}$. Lightweight regression and classification heads were attached with two ResNet blocks~\citep{he2016deep} and two $3\times3$ convolutions for one-to-one and one-to-many hybrid Hungarian matching. The models were trained and evaluated in an HPC environment using Nvidia A100 40GB GPUs. We provide training and inference runtime analysis for different approaches in~\ref{app2_runtime}.

We use $\tau=0.05$ as a weight term for the pixel distance in the Hungarian matching. In a hybrid Hungarian matching setup, $\beta=2$ was used for public datasets reported in Section~\ref{sec62}, while $\beta=6$ was used for the Patherea dataset, due to the increased resolution of the dataset. The influence of parameter $\beta$ is investigated in Section~\ref{sec64}. A class weighting term $\lambda_t$ in $\mathcal{L}_{cls}$ was set to 0.5 for the background class and 10 for all the foreground classes $t\in T$. Regression loss weight in $\mathcal{L}_{1v1}$ and $\mathcal{L}_{1vN}$ was set to $2e-3$, while $\lambda_{one2many}$ in a combined loss $\mathcal{L}$ was set to 0.5. Threshold on confidence $c_{thresh}^{t\in T}$ in inference was set to 0.9 for all classes $t\in T$ for Lizard~\citep{graham2021lizard} and BRCA-M2C~\citep{abousamra2021multi} datasets, while 0.5 was used for the Patherea dataset.


All models were initialized with ImageNet pre-trained weights for the backbone network, except where the pathology foundation model~\citep{nechaev2024hibou,chen2024towards} was explicitly employed. Training was conducted for 1000 epochs on public datasets (Section~\ref{sec62}) for both Patherea-P2P and FCRN++ methods. For experiments on the Patherea dataset (Section~\ref{sec63}), 100 epochs were used given the substantially larger dataset size, which would render extended training cycles computationally intractable for competing methods. All experiments employed a batch size of 16 and utilized the AdamW optimizer~\citep{loshchilov2018decoupled} with a learning rate of $1 \times 10^{-4}$, weight decay of $2 \times 10^{-3}$, and cosine annealing scheduling with a linear warm-up.

\subsection{Public Datasets}
\label{sec62}

\textbf{Lizard}: Lizard dataset was introduced in~\citep{graham2021lizard} and consists out of 6 different public datasets of colon cancer, where 291 images were extracted at x20 objective magnification. The Lizard dataset was primarily developed for nuclear instance segmentation and classification, but also provided cell center locations which can be utilized to develop and evaluate point-based approaches. The dataset was mostly automatically labeled, with a HoVer-Net~\citep{graham2019hover} trained on existing public data and used as an initial segmentation result. The segmentation results were later refined by a pathologist and a model was re-trained. Similarly, cell class refinement was augmented with pathologists-in-the-loop. This semi-automatic approach enabled the annotation of 495,179 cells of 6 different classes (epithelial, lymphocyte, plasma, neutrophil, eosinophil, connective).

The dataset was split into 3 folds and we followed the evaluation protocol used in ACFormer and PGT and used fold 3 for training, fold 2 for validation and fold 1 for testing. The ground truth radius $r$ was set to 6 pixels. We report the results in Table~\ref{tbl:lizard_results}.

\setlength{\tabcolsep}{1\tabcolsep}
\begin{table}[ht!]
\centering
\caption{Results on the Lizard dataset using 5x re-training strategy. Average F1 scores are reported. Standard deviation is also reported for $F1_{avg}$. $^\dag$ represents foundation model.}
\begin{tabular}{@{}llllllll@{}}
\toprule
            & $F1_{con}$ & $F1_{eos}$ & $F1_{epi}$ & $F1_{lym}$ & $F1_{neu}$ & $F1_{pla}$ & \textbf{$F1_{avg}$} \\ \toprule
MCSpatNet        & 0.659 & - & 0.800 & 0.705 & - & 0.437 & 0.434\small{$\;\pm 0.003$}  \\
ACFormer        & 0.723 & \textbf{0.565}  & \underline{0.816} & \textbf{0.771} & 0.405 & \textbf{0.580} & \underline{0.643}\small{$\;\pm 0.012$}  \\
PGT   & \underline{0.718} & 0.534  & \textbf{0.833} & \underline{0.757} & 0.283  & 0.538 & 0.611\small{$\;\pm 0.009$}  \\
FCRN++   & 0.573 & 0.019  & 0.720 &  0.664 & 0.048 & 0.289 & 0.386\small{$\;\pm 0.014$}\\ \bottomrule
\textbf{ours} (ViT)   & 0.665 & 0.452  & 0.782 &  0.734 & 0.366 & 0.500 & 0.583\small{$\;\pm 0.007$} \\
\textbf{ours} (ViT)$^\dag$   & 0.680 & 0.499  & 0.788 &  0.742 & \underline{0.444} & 0.524 & 0.613\small{$\;\pm 0.005$} \\
\textbf{ours} (CNN)   & \textbf{0.730} & \underline{0.543}  & \underline{0.816} &  \textbf{0.771} & \textbf{0.448} & \underline{0.569} & \textbf{0.646}\small{$\;\pm 0.007$} \\ \bottomrule
\end{tabular}
\label{tbl:lizard_results}
\end{table}

We report the results using the 5x re-training strategy to demonstrate real-world robustness and reproducibility. We report the average of the F1 scores across the 5 training runs. The official training code and recipes were used to re-train the models, except for the MCSpatNet and our re-implementation of the FCRN++ method, which did not report the results on the Lizard dataset. The checkpoint was selected using the best results on the validation set. We also report results using the officially provided pre-trained weights on Lizard for ACFormer and PGT in~\ref{app1_lizard}, where the improvements of our model are even more significant.

The best results are achieved using our proposed Patherea-P2P approach, using the ConvNext backbone. We hypothesize that the dataset of only 291 is too small for the ViT backbone to generalize and learn the inductive biases, despite ImageNet initialization. Another reason could be the low resolution of input images, scanned at 20x objective magnification and compressed, which basically removes most of the cell morphology information. We slightly improved ViT results by using a pathology foundation model$^\dag$ (Section~\ref{sec64}). We also noticed that density-based approaches MCSpatNet and FCRN++ performed significantly worse, mostly due to inability to detect eosinophil and neutrophil classes. These two classes are in low abundance (e.g., 3,604 and 4,824 labels in comparison with more than 100k for epithelial, lymphocyte and connective).

\textbf{BRCA-M2C}: This breast cancer dataset was introduced with the MCSpatNet~\citep{abousamra2021multi} and consists of 120 patches belonging to 113 patients, collected from TCGA. The dataset was collected at 20x objective magnification with inflammatory (lymphocyte), epithelial and stromal cells being labeled as point-based annotations. The ground truth radius $r=6$ was used for evaluation. We report results in Table~\ref{tbl:brca_results}.

\setlength{\tabcolsep}{1\tabcolsep}
\begin{table}[ht!]
\centering
\caption{Results on BRCA-M2C dataset using 5x re-training strategy. Average F1 scores are reported. Standard deviation is also reported for $F1_{avg}$. $^\dag$ represents foundation model.}
\begin{tabular}{@{}llllll@{}}
\toprule
            & $F1_{epi}$ & $F1_{lym}$ & $F1_{str}$ & \textbf{$F1_{avg}$} \\ \toprule
MCSpatNet        & 0.764 & 0.594 & 0.525 & 0.628\small{$\;\pm 0.009$}  \\
ACFormer        & 0.645 & 0.528  & 0.447 & 0.540\small{$\;\pm 0.084$}  \\
PGT   & \textbf{0.777} & \underline{0.625}  & \underline{0.524} & 0.642\small{$\;\pm 0.014$}  \\
FCRN++   & 0.725 & 0.579  & 0.472 &  0.592\small{$\;\pm 0.013$} \\ \bottomrule
\textbf{ours} (ViT)   & 0.753 & 0.629  & 0.502 &  0.628\small{$\;\pm 0.007$} \\
\textbf{ours} (ViT)$^\dag$   & 0.768 & \textbf{0.674}  & 0.509 &  \textbf{0.650}\small{$\;\pm 0.005$} \\
\textbf{ours} (CNN)   & \underline{0.772} & \underline{0.636}  & \textbf{0.525} &  \underline{0.644}\small{$\;\pm 0.007$}  \\ \bottomrule
\end{tabular}
\label{tbl:brca_results}
\end{table}

We report the results using the 5x re-training strategy and report the average F1 scores across the 5 training runs. Our proposed approach again achieved the best results using the ConvNext backbone, due to the small size of the BRCA-M2C dataset. The performance of the ViT-based backbone improved when the pathology foundation model$^\dag$ was used for pre-training. We also report results using the official weights for ACFormer and PGT in~\ref{app2_brca}.

\subsection{Patherea Dataset}
\label{sec63}

We evaluate the proposed Patherea-P2P method on our newly proposed Patherea dataset introduced in Section~\ref{sec4}. We split the Patherea dataset into 3 folds across different patients in a random manner and perform 3-fold cross-validation for all the approaches. The splits across the folds are released alongside the Patherea dataset. We used a fixed training schedule of 100 epochs for all the methods and report the average across the folds at 100 epochs. We report the results across different parts of the Patherea dataset - LNET in Table~\ref{tbl:lnet_results}, GNET in Table~\ref{tbl:gnet_results}, and Breast in Table~\ref{tbl:breast_results}.

Overall, the Patherea-P2P approach outperforms all of the competing approaches by a significant margin (e.g., 4-13\%) in comparison with the second best approach across different Patherea datasets. The improvement is more significant with less abundant classes of others positive and others negative classes. Interestingly, density-estimation-based approaches MCSpatNet and FCRN++ outperformed recently proposed DETR-based approaches.

\setlength{\tabcolsep}{1\tabcolsep}
\begin{table}[ht!]
\centering
\caption{Results on Patherea-LNET dataset. Average F1 scores are reported for 3-fold cross-validation.}
\begin{tabular}{@{}lllllll@{}}
\toprule
            & $F1_{pos}$ & $F1_{neg}$ & $F1_{othr\_pos}$ & $F1_{othr\_neg}$ & \textbf{$F1_{avg}$} \\ \toprule
MCSpatNet        & 0.790 & 0.734 & 0.190 & 0.446 & 0.540 \\
ACFormer        & 0.826 & 0.781  & - & - & 0.402 \\
PGT   & 0.789 & 0.758  & - & - & 0.387 \\
FCRN++   & 0.787 & 0.750  & 0.245 &  0.467 & 0.562 \\ \bottomrule
\textbf{ours} (ViT)   & \textbf{0.832} & \textbf{0.790}  & \underline{0.273} &  \textbf{0.572} & \underline{0.617} \\
\textbf{ours} (CNN)   & \underline{0.828} & \underline{0.789}  & \textbf{0.318} & \underline{0.559} & \textbf{0.624} \\ \bottomrule
\end{tabular}
\label{tbl:lnet_results}
\end{table}

\setlength{\tabcolsep}{1\tabcolsep}
\begin{table}[ht!]
\centering
\caption{Results on Patherea-GNET dataset. Average F1 scores are reported for 3-fold cross-validation.}
\begin{tabular}{@{}lllllll@{}}
\toprule
            & $F1_{pos}$ & $F1_{neg}$ & $F1_{othr\_pos}$ & $F1_{othr\_neg}$ & \textbf{$F1_{avg}$} \\ \toprule
MCSpatNet        & 0.612 & 0.825 & 0.127 & 0.579 & 0.536  \\
ACFormer        & 0.584 & 0.827  & - & - & 0.353 \\
PGT   & 0.598 & 0.825  & - & - & 0.356 \\
FCRN++   & 0.746 & \underline{0.842}  & - &  0.629 & 0.555 \\ \bottomrule
\textbf{ours} (ViT)   & \textbf{0.774} & \textbf{0.852}  & \textbf{0.241} &  \textbf{0.658} & \textbf{0.631} \\
\textbf{ours} (CNN)   & \underline{0.756} & \underline{0.842}  & \underline{0.214} & \underline{0.649} & \underline{0.615} \\ \bottomrule
\end{tabular}
\label{tbl:gnet_results}
\end{table}

Patherea-P2P method based on the ViT backbone consistently outperformed the ConvNext backbone. This implies that our proposed approach effectively utilizes the additional scale of the data, in comparison with existing public datasets used in Section~\ref{sec62}, where performance is largely saturated. Patherea dataset can thus serve as baseline dataset to benchmark new proposed approaches. Additionally, the others positive and others negative classes present a significant challenge due to their visual similarity with positive and negative tumor cells. We hypothesize that the performance on those two classes could be further improved with approaches that can more effectively handle imbalanced datasets.

\setlength{\tabcolsep}{1\tabcolsep}
\begin{table}[ht!]
\centering
\caption{Results on Patherea-Breast dataset. Average F1 scores are reported for 3-fold cross-validation for both pathologists (P1/P2).}
\centerline{
\begin{tabular}{@{}lllllll@{}}
\toprule
            & $F1_{pos}$ & $F1_{neg}$ & $F1_{othr\_pos}$ & $F1_{othr\_neg}$ & \textbf{$F1_{avg}$} \\ \toprule
MCSpatNet        & 0.771/0.690 & 0.680/0.696 & 0.315/- & \underline{0.606}/0.511 & 0.593/0.474  \\
ACFormer        & 0.761/0.717 & 0.657/0.668  & -/- & - /-& 0.378 \\
PGT   & \underline{0.794}/\underline{0.770} & 0.713/0.742  & -/- & -/- & 0.377/0.378 \\
FCRN++   & 0.785/0.768 & \underline{0.736}/\underline{0.762}  & \underline{0.326}/0.025 &  0.586/0.533 & \underline{0.608}/\underline{0.522} \\ \bottomrule
\textbf{ours} (ViT)   & \textbf{0.804}/\textbf{0.776} & \textbf{0.748}/\textbf{0.763}  & \textbf{0.341}/\underline{0.143} &  \textbf{0.640}/\textbf{0.585} & \textbf{0.633}/\textbf{0.567} \\
\textbf{ours} (CNN)   & 0.752/0.714 & 0.698/0.722  & 0.319/\textbf{0.169} & \underline{0.607}/0.531 & 0.594/\underline{0.534} \\ \bottomrule
\end{tabular}
}
\label{tbl:breast_results}
\end{table}

\subsection{Qualitative Results}
\label{sec644}

\begin{figure}[!htbp]
\centering
\includegraphics[width=\textwidth,height=\textheight,keepaspectratio]{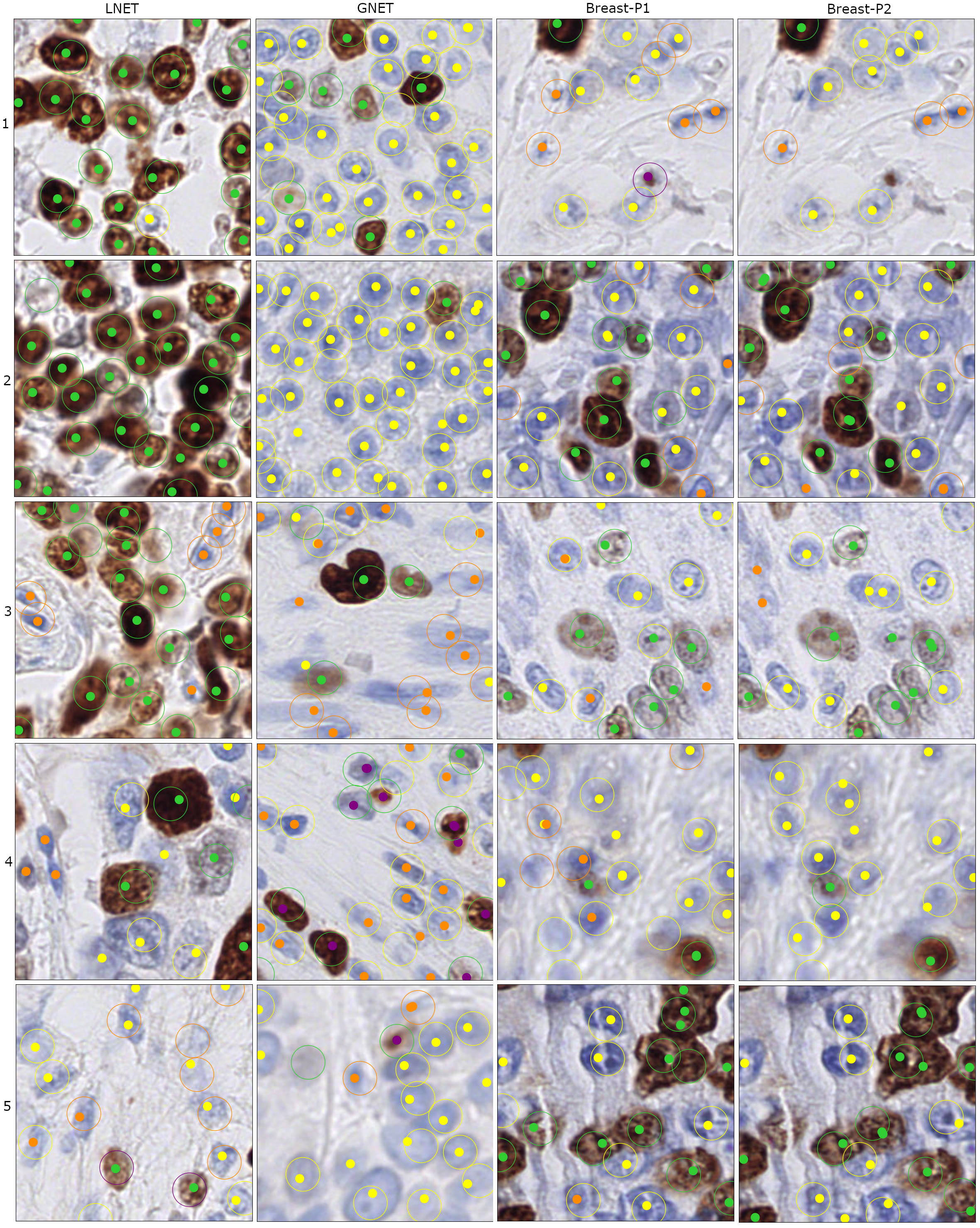}
\caption{Patherea-P2P (ViT) qualitative results on different Patherea datasets. Example detections are displayed as filled circles, color-coded for cell type ($\greendot\;pos$, $\yellowdot\; neg$, $\purpledot\; othr\_pos$, $\orangedot\; othr\_neg$). Ground-truth circular region is color coded (border) with the ground-truth cell type. Best viewed in an online version.}
\label{fig_collage_patherea}
\end{figure}

Qualitative results for the Patherea-P2P model are presented in Figure~\ref{fig_collage_patherea}. First, we notice that proposal candidates are successfully filtered down with the Hungarian one-to-one matching and confidence thresholding (0.5), such that multiple detections occur rarely. This also holds true in most of the extreme cases, where cell size is significantly larger (e.g., LNET-4, GNET-3). Cells are successfully detected in dense regions (e.g., LNET-2, GNET-2), as well as when cell morphology is less apparent (e.g., GNET-5, Breast-4). There are some multiple detections present when cells are significantly overlapped and clustered (e.g., Breast-5). One could utilize the non-maxima suppression to fine-tune further the predictions, which would result in the introduction of the post-processing step and break the fully end-to-end learning-based paradigm. Qualitative comparative analysis against other approaches is presented in Figure~\ref{fig_collage_patherea_2}.

Labeling cells directly on whole slide images (WSIs), as done in real-world diagnostic workflows, poses a substantially greater challenge than annotating isolated image patches. Variability arises not only from missed annotations but also from subjective differences in the thresholds applied by individual pathologists when labeling specific cell types. For instance, we observe model-predicted detections that were not annotated by experts (e.g., LNET-4, GNET-3), suggesting possible under-labeling or inter-observer variability. In the Breast-4 sample, the model appears to align closely with the labeling behavior of the specific pathologist, highlighting the model’s ability to learn annotator-specific patterns. Notably, breast samples were independently labeled twice by different pathologists, and their annotations often diverge in borderline categories — particularly between others positive and others negative — emphasizing the subjectivity involved in such classifications.


In Figures~\ref{fig_collage_lizard} and~\ref{fig_collage_brca_m2c} we present qualitative results for Lizard~\citep{graham2021lizard} and BRCA-M2C~\citep{abousamra2021multi} datasets. We notice that the resolution is significantly lower with significantly less cell morphology visible to differentiate different cell types. We notice a confident detection of well-representative classes and similarly some spurious detections in regions where no labels are present.

\subsection{Ablation Studies}
\label{sec64}

\textbf{Hybrid Hungarian Matching}: In Table~\ref{tbl:hybrid_hungarian} we evaluate the influence of the parameter $\beta$ which is used to select the number of candidates each ground truth target is matched against for an additional one-to-many loss in Patherea-P2P, as presented in Section~\ref{subsec_hp2p}. We demonstrate the effectiveness of the proposed hybrid matching, with relative performance gains of up to 7\% for sufficiently represented classes and up to 24\% for less abundant classes.

\setlength{\tabcolsep}{1\tabcolsep}
\begin{table}[ht!]
\centering
\caption{The influence of the number of matching proposal candidates in training - $\beta$ for Patherea-LNET dataset. Reported F1 scores are averaged across the 3 folds. positive$_{othr}$ results are reported at 50 and 100 epochs. Relative improvement $\Delta$ is also reported against the default setup ($\beta = 1$).}
\begin{tabular}{@{}l|l|lll|ll@{}}
\toprule
     $\beta$       & 1 & 2 & 4 & 6 & $\Delta[\%]$ \\ \toprule
positive        & 0.799 & 0.820 & 0.828 & \textbf{0.832} & 4.1  \\
negative        & 0.744 & 0.772  & 0.788 & \textbf{0.790} & 6.2  \\
positive$_{othr100}$   & 0.292 & \textbf{0.332}  & 0.287 & 0.273 & 13.7  \\
positive$_{othr50}$   & 0.268 & 0.311  & \textbf{0.332} & 0.319 & 23.9  \\
negative$_{othr}$   & 0.534 & 0.563  & 0.567 &  \textbf{0.572} & 7.1 \\ \bottomrule
\end{tabular}
\label{tbl:hybrid_hungarian}
\end{table}

We observe that the performance on the Patherea-LNET dataset improves with increasing values of $\beta$, with performance gains beginning to plateau between $\beta=4$ and $\beta=6$. Accordingly, we set $\beta=6$ as the default for the Patherea dataset in Section~\ref{sec63}. In contrast, for the Lizard and BRCA-M2C datasets, a lower value of $\beta=2$ was used in Section~\ref{sec62}, reflecting the lower resolution associated with the 20$\times$ objective magnification used in those datasets.

We also notice an overfitting against the extremely imbalanced positive$_{othr}$ class when evaluated at 50 and 100 epochs. The positive class starts to dominate and reduces the performance on the positive$_{othr}$ class with longer training.

\textbf{Foundation Model}: In Table~\ref{tbl:foundation} we report the average F1 performance on Lizard~\citep{graham2021lizard}, BRCA-M2C~\citep{abousamra2021multi} and Patherea-LNET datasets when using different model training strategies. For Lizard and BRCA-M2C we report the $F1_{avg}$ across different classes with a 5x re-training strategy, directly comparable with results reported in Tables~\ref{tbl:lizard_results} and~\ref{tbl:brca_results}. Similarly to Table~\ref{tbl:lnet_results}, we report $F1_{avg}$ across the 3 folds for the Patherea-LNET dataset.

\setlength{\tabcolsep}{1\tabcolsep}
\begin{table}[ht!]
\centering
\caption{Average F1 performance when training from scratch, fine-tuning from ImageNet or using the pathology foundation model Hibou.}
\begin{tabular}{@{}lllll@{}}
\toprule
            & Lizard & BRCA-M2C  & LNET     \\ \toprule
Scratch     & 0.459  & 0.471     & 0.576    \\
ImageNet    & 0.583  & 0.628     & \textbf{0.617}    \\
Hibou       & \textbf{0.613}  & \textbf{0.650}     & \textbf{0.617}    \\ \bottomrule
\end{tabular}
\label{tbl:foundation}
\end{table}

We evaluated three different training strategies for training Patherea-P2P using the ViT backbone. With the \textit{Scratch} strategy, we trained the Patherea-P2P model from scratch (backbone, ViT-Adapter, heads). The 2nd approach, fine-tuning from \textit{ImageNet} initialized ViT backbone represents the default approach used to report results in Sections~\ref{sec62} and~\ref{sec63}. Lastly, we utilized the recently released \textit{Hibou} family of open-source foundation models for pathology~\citep{nechaev2024hibou}. For \textit{ImageNet}, we fine-tuned the whole Patherea-P2P architecture, while for \textit{Hibou}, we only fine-tuned ViT-Adapter~\citep{chen2023vision} and heads and kept the ViT backbone frozen.

We notice that training from scratch significantly reduces the performance on the Lizard and BRCA-M2C datasets, while the drop on the Patherea-LNET dataset is smaller. We hypothesize that this is due to the significantly smaller size of the Lizard and BRCA-M2C datasets in terms of the actual number of training samples. We also notice a 3-5\% improvement over ImageNet when using a foundation model on Lizard and BRCA-M2C datasets. We were able to improve upon ConvNext-based Patherea-P2P results on BRCA-M2C, reported in Table~\ref{tbl:brca_results}, when using a foundation model. No improvement was observed on the Patherea-LNET dataset when using a pre-trained \textit{Hibou} foundation model.

We hypothesize that this is due to the training data used in \textit{Hibou}, where around 80\% of the training samples are H\&E stains, with the other 20\% represented by other stains, which are not specified. This is common with the current publicly available foundation models, as immunohistochemistry samples are not widely available. We additionally confirm the improved results on Lizard and BRCA-M2C using the UNI foundation model~\citep{chen2024towards} in Table~\ref{tbl:foundation_uni}. We expect to observe similar gains on the Patherea dataset with foundation models, specifically trained on Ki-67 samples.

In Table~\ref{tbl:lizard_results_cellvit}, we also compared our Patherea-P2P approach against the recently proposed instance segmentation-based approach CellViT$^{++}$\citep{cellvit++}, which utilizes foundation models to reduce the need for costly segmentation mask labels.

\section{Conclusion}
\label{sec7}

In this work, we presented Patherea, a comprehensive framework for developing clinically applicable AI models for point-based cell detection and classification. Diagnostic workflows involving immunohistochemistry, such as Ki-67 proliferation index estimation, remain labor-intensive and subject to interobserver variability. Automating these tasks is of critical importance, yet it is hindered by the scarcity of large, high-quality annotated datasets and the limited capacity of existing methods to effectively utilize point-based annotations—the most efficient annotation modality in clinical practice.

To address this gap, we curated the largest fully manually annotated dataset with point-based labels for Ki-67 estimation in Neuroendocrine tumors (NETs) and breast cancer. Importantly, the annotation protocol was designed to closely mirror real-world clinical workflows: pathologists performed cell annotations while having full access to the entire whole slide image (WSI), rather than being constrained to isolated image patches. This approach ensures that annotations reflect the contextual decision-making process used in routine diagnostics. The resulting dataset thus provides both a realistic benchmark for evaluating automated methods and a high-quality resource for training models that are aligned with actual clinical practice.

We introduced a novel approach for cell detection and classification that learns directly from point annotations without requiring intermediate representations such as bounding boxes or segmentation masks. Our proposed method includes an improved target-candidate association strategy that considers multiple high-quality candidates for each annotated point, thereby enhancing robustness and precision. The resulting model, Patherea-P2P, achieves state-of-the-art performance on existing large-scale public benchmarks and demonstrates superior generalization on the newly introduced Patherea dataset, which poses substantial challenges due to its clinical complexity.

Additionally, we demonstrated that the proposed architecture is well-suited for leveraging large-scale unlabeled data through foundation model pre-training. This enables effective transfer learning and fine-tuning with limited labeled data, which is critical for real-world clinical applications. We also revisited the evaluation protocols commonly used in point-based detection and classification tasks, proposing a revised and more accurate scoring scheme to ensure reliable and meaningful comparisons across methods.

While the focus of this study is the technical development of point-based cell detection and classification methods, our ultimate goal is to support accurate and reproducible estimation of clinically relevant biomarkers, such as the Ki-67 index. As part of future work, we plan to apply the proposed framework to large cohorts of Ki-67 stained samples annotated with slide-level index scores from multiple pathologists. This will enable rigorous evaluation of the clinical validity of our approach and further establish its utility in diagnostic pathology.

\section{Acknowledgment}
\label{sec7}

The research was partially co-financed by funds from the Slovenian Research and Innovation Agency (ARIS), allocated to the research program groups P3-0054 (Pathology and Molecular Genetics) and P2-0214 (Computer Vision). The authors gratefully acknowledge the HPC RIVR consortium and EuroHPC JU for funding this research by providing computing resources of the HPC system Vega at the Institute of Information Science, Maribor, Slovenia.

\newpage
\appendix

\section{Patherea Dataset Annotation}
\label{sec:supp_figures}

The following cell types were annotated by pathologists in the Ki67-stained tissue sections:

\begin{itemize}
\item \textbf{Positive tumor cells:} tumor cells that stain positive for Ki67, indicating proliferating cells.
\item \textbf{Negative tumor cells:} tumor cells that do not stain positive for Ki67, indicating resting (quiescent) cells.
\item \textbf{Other positive cells:} Normal cells that stain positive for Ki67, such as endothelial cells, inflammatory cells, and mucosal epithelial cells, which are in proliferation phases.
\item \textbf{Other negative cells:} Non-tumorous cells that do not stain positive for Ki67, for example, endothelial cells, inflammatory cells, non-neoplastic ducts and mucosal epithelial cells, which are not in proliferating phases.
\end{itemize}

\begin{figure}[H]
\centering
\includegraphics[width=\textwidth,height=\textheight,keepaspectratio]{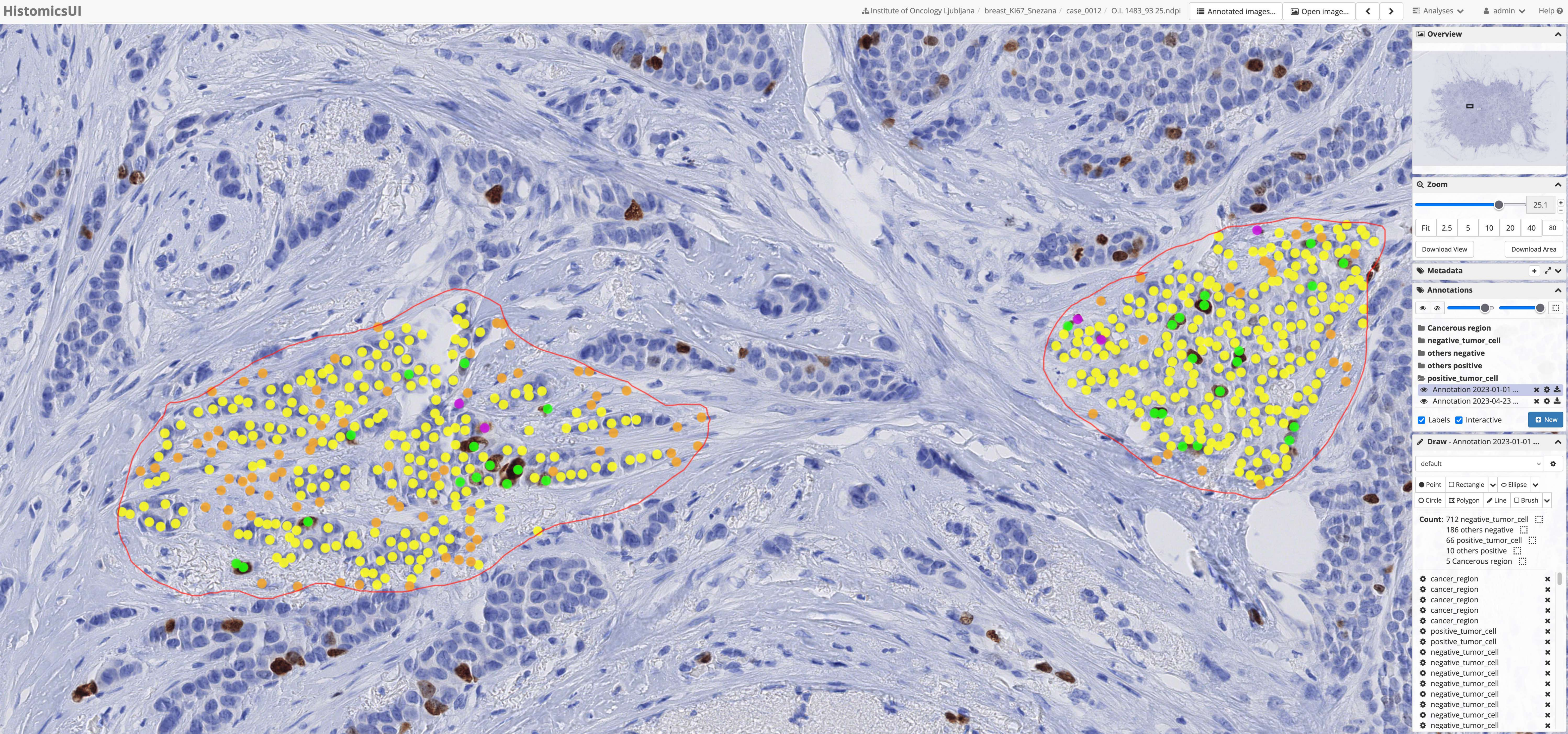}
\caption{An example of the labeled WSI from the Digital Slide Archive~\citep{gutman2017digital}. The pathologists selected the polygon regions (red) and labeled all the cells and their types using point-based annotations. Different colors represent the four different cell types color-coded as $\greendot\;pos$, $\yellowdot\; neg$, $\purpledot\; othr\_pos$, $\orangedot\; othr\_neg$.}
\label{fig:sup_platform}
\end{figure}

\section{Additional Results}
\label{app1}

\subsection{Lizard}
\label{app1_lizard}

In Table~\ref{tbl:lizard_results_2} we report the results on the Lizard dataset~\citep{graham2021lizard} by using officially provided weights to initialize the ACFormer~\citep{huang2023affine} and PGT~\citep{huang2023prompt} methods, without the need for re-training. No official training weights were provided for MCSpatNet~\citep{abousamra2021multi} and FCRN++.


We observe a slight performance drop when using official weights. We report the best Patherea-P2P and FCRN++ results from 100 training runs with different random seeds, providing an upper bound for the Lizard dataset. Our efficient Patherea-P2P implementation enables this extensive evaluation, requiring only ~2 hours of training per-run on a single GPU, compared to 2 days on 4 GPUs for MCSPatNet, ACFormer, and PGT (\ref{app2_runtime}).

\setlength{\tabcolsep}{1\tabcolsep}
\begin{table}[ht!]
\centering
\caption{Results on Lizard dataset using official pre-trained weights for PGT and ACFormer. Best out of 5 training runs is reported for MCSpatNet. Best out of 100 training runs using different random seeds is reported for FCRN++ and Patherea-P2P.}
\begin{tabular}{@{}llllllll@{}}
\toprule
            & $F1_{con}$ & $F1_{eos}$ & $F1_{epi}$ & $F1_{lym}$ & $F1_{neu}$ & $F1_{pla}$ & \textbf{$F1_{avg}$} \\ \toprule
MCSpatNet        & 0.661 & 0 & 0.814 & 0.712 & 0 & 0.442 & 0.438  \\
ACFormer        & \underline{0.714} & 0.506  & \underline{0.819} & 0.740 & 0.330 & \underline{0.561} & 0.612  \\
PGT   & 0.712 & \underline{0.536}  & 0.810 & \underline{0.742} & \underline{0.395}  & 0.515 & \underline{0.618}  \\
FCRN++   & 0.594 & 0.017  & 0.733 &  0.676 & 0.055 & 0.309 & 0.397\\ \bottomrule
\textbf{ours} (CNN)   & \textbf{0.722} & \textbf{0.570}  & \textbf{0.815} &  \textbf{0.780} & \textbf{0.445} & \textbf{0.585} & \textbf{0.653} \\ \bottomrule
\end{tabular}
\label{tbl:lizard_results_2}
\end{table}

\subsection{BRCA-M2C}
\label{app2_brca}

We notice that MCSpatNet~\citep{abousamra2021multi} performs much better with the official weights in comparison with 5x re-training (reported in Table~\ref{tbl:brca_results}) using the official code and training recipes. We report the best obtained Patherea-P2P and FCRN++ results after 100 training runs using different random seeds. We notice that we are able to reach the official performance of the MCSpatNet with one of the random seeds, which suggests that the best-performing MCSpatNet weights were selected for release. Similar improvement is noticeable for ACFormer~\citep{huang2023affine}.

\setlength{\tabcolsep}{1\tabcolsep}
\begin{table}[ht!]
\centering
\caption{Results on BRCA-M2C dataset using official pre-trained weights for MCSpatNet, PGT and ACFormer. Best out of 100 training runs using different random seeds is reported for FCRN++ and Patherea-P2P.}
\begin{tabular}{@{}llllll@{}}
\toprule
            & $F1_{epi}$ & $F1_{lym}$ & $F1_{str}$ & \textbf{$F1_{avg}$} \\ \toprule
MCSpatNet        & \textbf{0.793} & 0.649 & \textbf{0.548} & \textbf{0.663}  \\
ACFormer        & 0.741 & 0.618  & 0.462 & 0.607  \\
PGT   & \textbf{0.792} & 0.624  & 0.531 & 0.649  \\
FCRN++   & 0.777 & 0.583  & 0.489 &  0.616 \\ \bottomrule
\textbf{ours} (CNN)   & \underline{0.786} & \textbf{0.659}  & \underline{0.538} &  \underline{0.661} \\ \bottomrule
\end{tabular}
\label{tbl:brca_results_2}
\end{table}

\subsection{BCData}
\label{app2_bcdata}

We report results on BCData~\citep{huang2020bcdata} in Table~\ref{tbl:bcdata}. We directly report F1 scores from the original publication for competing methods SC-CNN~\citep{sirinukunwattana2016locality}, CSRNet~\citep{li2018csrnet} and U-CSRNet~\citep{huang2020bcdata}. The official implementation code was not released. No details (or code) were provided on F1 score calculation. FCRN re-implementation~\citep{xie2018efficient} and Patherea-P2P results are reported using our F1 benchmarking code (Section~\ref{sec_5}).

\setlength{\tabcolsep}{1\tabcolsep}
\begin{table}[ht!]
\centering
\caption{Results on BCData dataset. Best out of 100 training runs using different random seeds is reported for FCRN++ and Patherea-P2P.}
\begin{tabular}{@{}llll@{}}
\toprule
            & $F1_{epi}$ & $F1_{lym}$ & \textbf{$F1_{avg}$} \\ \toprule
SC-CNN        & 0.798 & 0.778 & 0.788  \\
CSRNet        & 0.829 & 0.814 & 0.822  \\
U-CSRNet   & \textbf{0.863} & \underline{0.852} & \underline{0.857}  \\
FCRN++   & 0.844 & 0.804 & 0.824 \\ \bottomrule
\textbf{ours} (CNN)   & \underline{0.862} & \textbf{0.854} & \textbf{0.858} \\ \bottomrule
\end{tabular}
\label{tbl:bcdata}
\end{table}

FCRN++ implementation is very similar to U-CSRNet~\citep{huang2020bcdata}, but achieves a lower performance. No implementation code or benchmarking details are available for U-CSRNet to verify the results.

\subsection{Hungarian Matching - Evaluation}
\label{app2_hungarian}

In Table~\ref{tbl:lnet_results_hung_dist} we report Patherea-LNET evaluation results using the Hungarian algorithm on the distance matrix, as opposed to the Boolean thresholded matrix (Section~\ref{sec_5}). Note that we only repeated the evaluation step on the same model predictions, as reported in Table~\ref{tbl:lnet_results}.

\setlength{\tabcolsep}{1\tabcolsep}
\begin{table}[ht!]
\centering
\caption{Results on Patherea-LNET dataset by using the Hungarian algorithm on the distance matrix. Average F1 scores are reported for 3-fold cross-validation.}
\begin{tabular}{@{}llllll|ll@{}}
\toprule
            & $F1_{pos}$ & $F1_{neg}$ & $F1_{othr\_pos}$ & $F1_{othr\_neg}$ & \textbf{$F1_{avg}$} & $\Delta[\%]$ \\ \toprule
MCSpatNet        & 0.753 & 0.706 & 0.189 & 0.430 & 0.520 & -3.8  \\
ACFormer        & 0.794 & 0.754  & - & - & 0.387 & -3.9 \\
PGT   & 0.766 & 0.735  & - & - & 0.375 & -3.2\\
FCRN++   & 0.779 & 0.745  & 0.203 &  0.451 & 0.545 & -3.1 \\ \bottomrule
\textbf{ours} (ViT)   & \textbf{0.800} & \textbf{0.772}  & \underline{0.259} &  \textbf{0.559} & \textbf{0.598}  & -3.2\\
\textbf{ours} (CNN)   & \textbf{0.800} & \underline{0.767}  & \textbf{0.280} & \underline{0.545} & \textbf{0.598} & -4.3 \\ \bottomrule
\end{tabular}
\label{tbl:lnet_results_hung_dist}
\end{table}

The results support the claims from Section~\ref{sec_5} that the global objective in the Hungarian algorithm underestimates F1 scores due to some portion of wrongly matched predictions. The relative performance/ranking of the different approaches has not changed, as the error is roughly equally distributed (3-4\%).

\subsection{UNI Foundation Model}
\label{app2_foundation}

In Table~\ref{tbl:foundation_uni} we report the results of using UNI pretrained foundation model~\citep{chen2024towards}. We used the same evaluation setup as for the results presented in Table~\ref{tbl:foundation}. ViT-L backbone~\citep{dosovitskiy2021an} was used for Scratch, ImageNet, and UNI, in comparison to the ViT-B backbone used with Hibou experiments in Table~\ref{tbl:foundation} - as UNI pre-trained weights are not available for ViT-B. UNI family of models were not trained on any of the standard open datasets that are often used in labeled datasets, such as Lizard~\citep{graham2021lizard} and BRCA-M2C~\citep{abousamra2021multi}. In comparison to Hibou~\citep{nechaev2024hibou}, only H\&E slides were included in the UNI training dataset.

\setlength{\tabcolsep}{1\tabcolsep}
\begin{table}[ht!]
\centering
\caption{Average F1 performance when training from scratch, fine-tuning from ImageNet or using the pathology foundation model UNI.}
\begin{tabular}{@{}lllll@{}}
\toprule
            & Lizard & BRCA-M2C  & LNET     \\ \toprule
Scratch     & 0.427  & 0.507     & 0.559    \\
ImageNet    & 0.562  & 0.598     & \textbf{0.612}    \\
UNI       & \textbf{0.624}  & \textbf{0.661}     & \textbf{0.610}    \\ \bottomrule
\end{tabular}
\label{tbl:foundation_uni}
\end{table}

The results demonstrate a similar behaviour to Hibou results presented in Table~\ref{tbl:foundation}. The results demonstrate that the boost on the Lizard and BRCA-M2C datasets is not from the TCGA data leakage, but rather an actual boost from the pre-trained model that our proposed method can effectively utilize.

\subsection{Segmentation Model Comparison}
\label{app2_Segmentation}

In Table~\ref{tbl:lizard_results_cellvit} we compare our \textit{Hibou} foundation model approach from Table~\ref{tbl:lizard_results} against the CellViT$^{++}$\citep{cellvit++} instance segmentation approach.

\setlength{\tabcolsep}{1\tabcolsep}
\begin{table}[ht!]
\centering
\caption{Comparison against the CellViT$^{++}$ instance segmentation approach. $^\dag$ represents a foundation model.}
\begin{tabular}{@{}llllllll@{}}
\toprule
            & $F1_{con}$ & $F1_{eos}$ & $F1_{epi}$ & $F1_{lym}$ & $F1_{neu}$ & $F1_{pla}$ & \textbf{$F1_{avg}$} \\ \toprule
CellViT$^{++}$\small{(SAM-H)}  & 0.622 & 0.327  & 0.787 &  0.678 & 0.331 & 0.422 & 0.528\\ \bottomrule
\textbf{ours} (ViT)$^\dag$   & \textbf{0.680} & \textbf{0.499}  & \textbf{0.788} &  \textbf{0.742} & \textbf{0.444} & \textbf{0.524} & \textbf{0.613}\small{$\;\pm 0.005$} \\ \bottomrule
\end{tabular}
\label{tbl:lizard_results_cellvit}
\end{table}

Our approach performed significantly better, despite being only trained on point-based annotations. In both cases, the backbone was frozen. For CellViT$^{++}$, only a lightweight cell classification module was re-trained, while only ViT-Adapter\-~\citep{chen2023vision} layers and prediction heads were trained for Patherea-P2P. Pre-trained segmentation backbone of CellViT$^{++}$ improves the performance on rare classes (i.e., eosinophil, neutrophil) against some of the competing approaches in Table~\ref{tbl:lizard_results}.

\subsection{Runtime Analysis}
\label{app2_runtime}

We report an approximate training time per-fold on Patherea-LNET dataset in Table~\ref{tbl:lnet_runtime}. The training time of our proposed Patherea-P2P method is multi-fold faster, regardless of the chosen backbone. Note that we didn't include the pre-processing time needed to create intermediate representations for MCSpatNet~\citep{abousamra2021multi} and FCRN++. We also only report the prompt fine-tuning training time for the PGT~\citep{huang2023prompt} method.

The average inference time per image sample ($224 \times 224$) is also reported. We used one of the Patherea-LNET test folds to compute the average inference time with a batch size set to 1 for all the approaches using a single Nvidia A100 40GB GPU. The proposed ViT-based approach is significantly faster in comparison with DETR-based approaches (ACFormer~\citep{huang2023affine}, PGT~\citep{huang2023prompt}), while the CNN backbone offers similar performance to density-based approaches (MCSpatNet~\citep{abousamra2021multi}, FCRN++.

\setlength{\tabcolsep}{1\tabcolsep}
\begin{table}[ht!]
\centering
\caption{Training (per fold - $t_{fold}$) and inference (per $224 \times 224$ sample - $t_{sample}$) runtime results on Patherea-LNET dataset.}
\begin{tabular}{@{}lll@{}}
\toprule
            & $t_{fold}$ & $t_{sample}$ [ms]\\ \toprule
MCSpatNet   & 1d18h   & \textbf{5}  \\
ACFormer    & 7d   & 170 \\
PGT     & 7d4h   & 45\\
FCRN++    & 4h  & 9 \\ \bottomrule
\textbf{ours} (ViT) & 2.1h    & 21 \\
\textbf{ours} (CNN) & \textbf{1.5h}   & \underline{13} \\ \bottomrule
\end{tabular}
\label{tbl:lnet_runtime}
\end{table}


\subsection{Qualitative Results}
\label{app3_qualitative}

\begin{figure}[!htbp]
\centering
\includegraphics[width=\textwidth,height=\textheight,keepaspectratio]{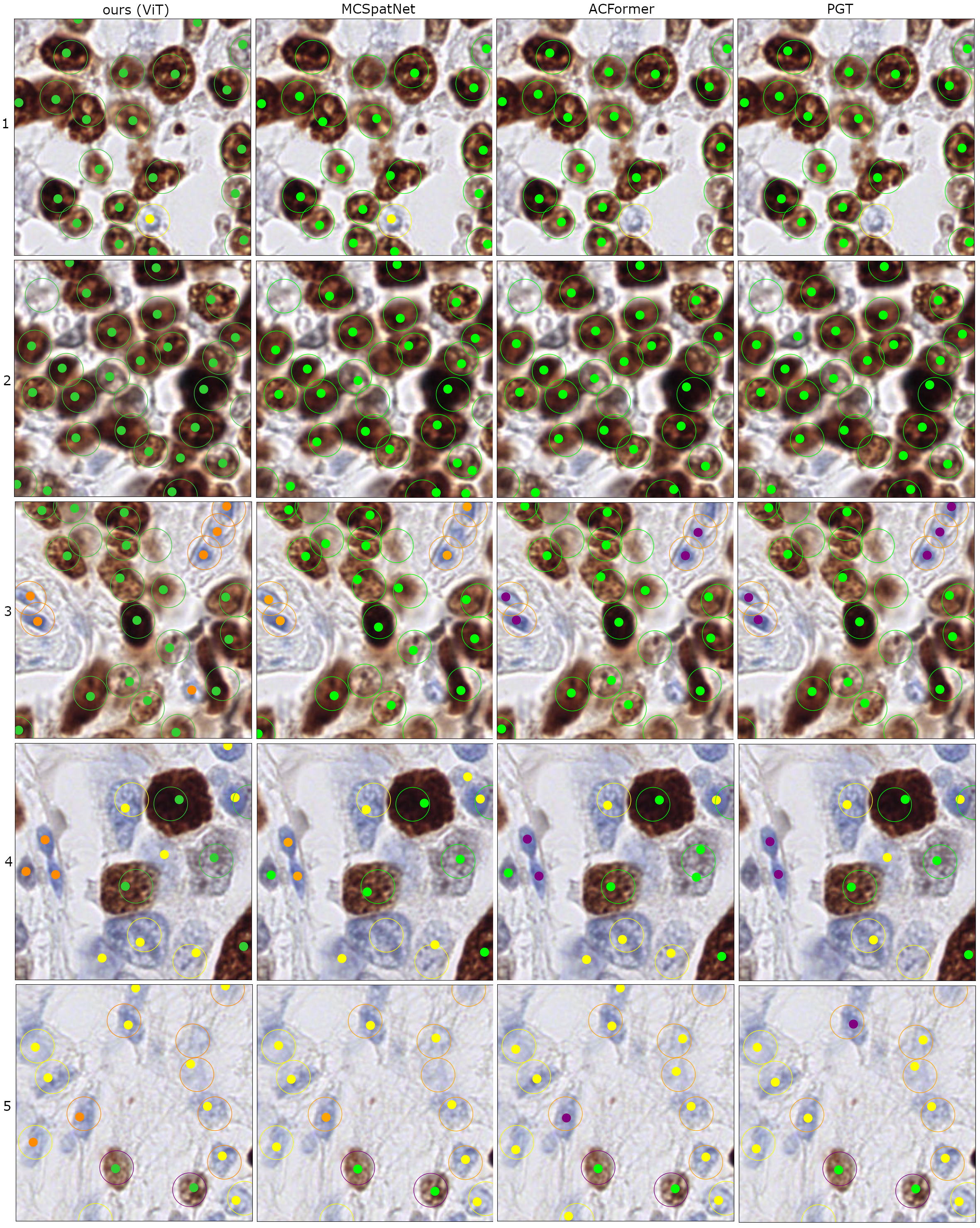}
\caption{Qualitative results of different approaches on Patherea-LNET dataset. Example detections are displayed as filled circles, color-coded for cell type ($\greendot\;pos$, $\yellowdot\; neg$, $\purpledot\; othr\_pos$, $\orangedot\; othr\_neg$). Ground-truth circular region is color coded (border) with the ground-truth cell type. Best viewed in an online version.}
\label{fig_collage_patherea_2}
\end{figure}

\begin{figure}[!htbp]
\centering
\includegraphics[width=1\textwidth,height=1\textheight,keepaspectratio]{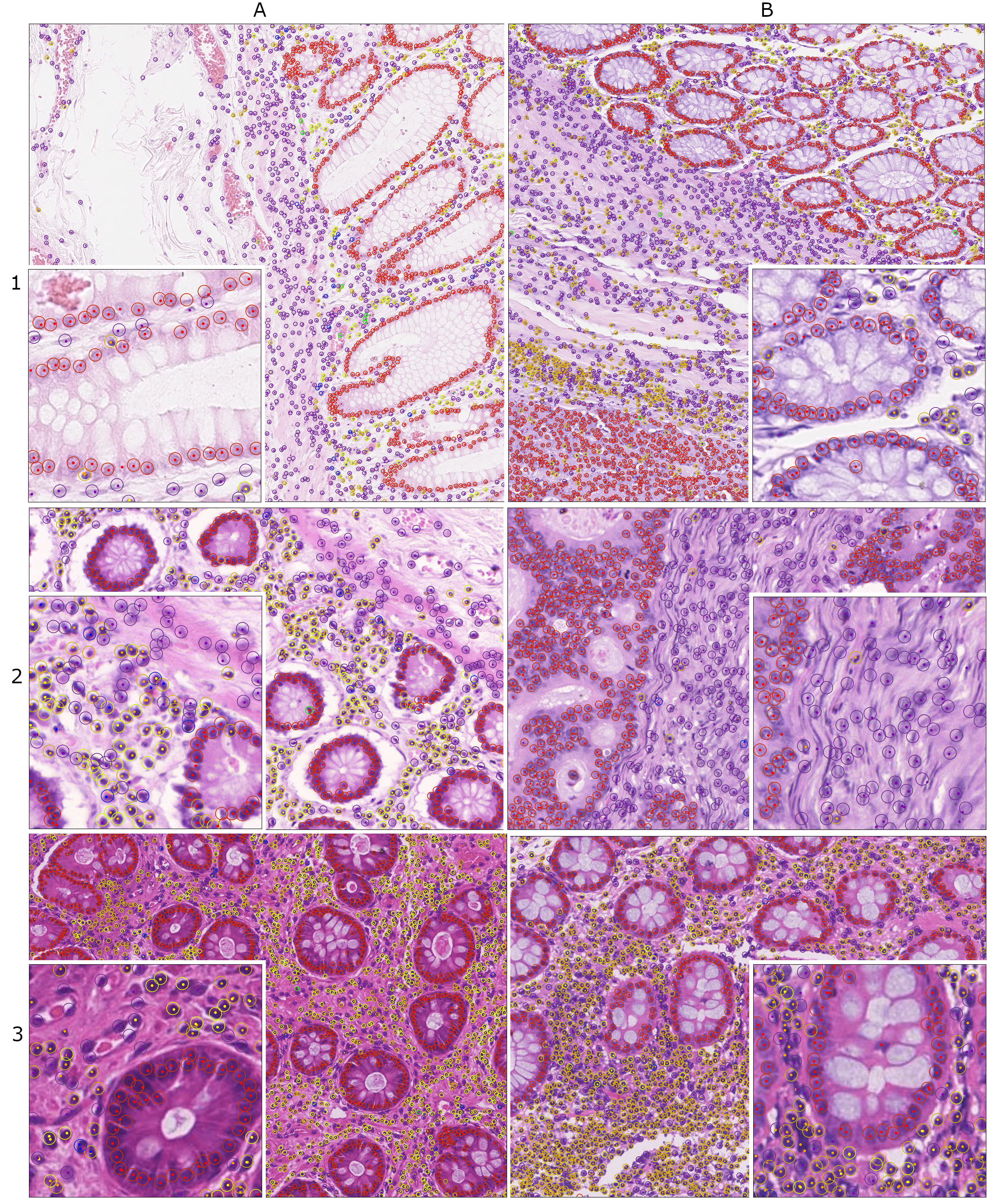}
\caption{Patherea-P2P (ConvNext) qualitative results on Lizard dataset. Example detections are displayed as filled circles, color-coded for cell type ($\greendot\; neutrophil$, $\reddot\; epithelial$, $\orangedot\; lymphocyte$, $\yellowdot\; plasma$, $\bluedot\; eosinophil$, $\purpledot\; connective$). Ground-truth circular region is color coded (border) with the ground-truth cell type. Best viewed in an online version with zoom.}
\label{fig_collage_lizard}
\end{figure}

\begin{figure}[!htbp]
\centering
\includegraphics[width=1\textwidth,height=0.9\textheight,keepaspectratio]{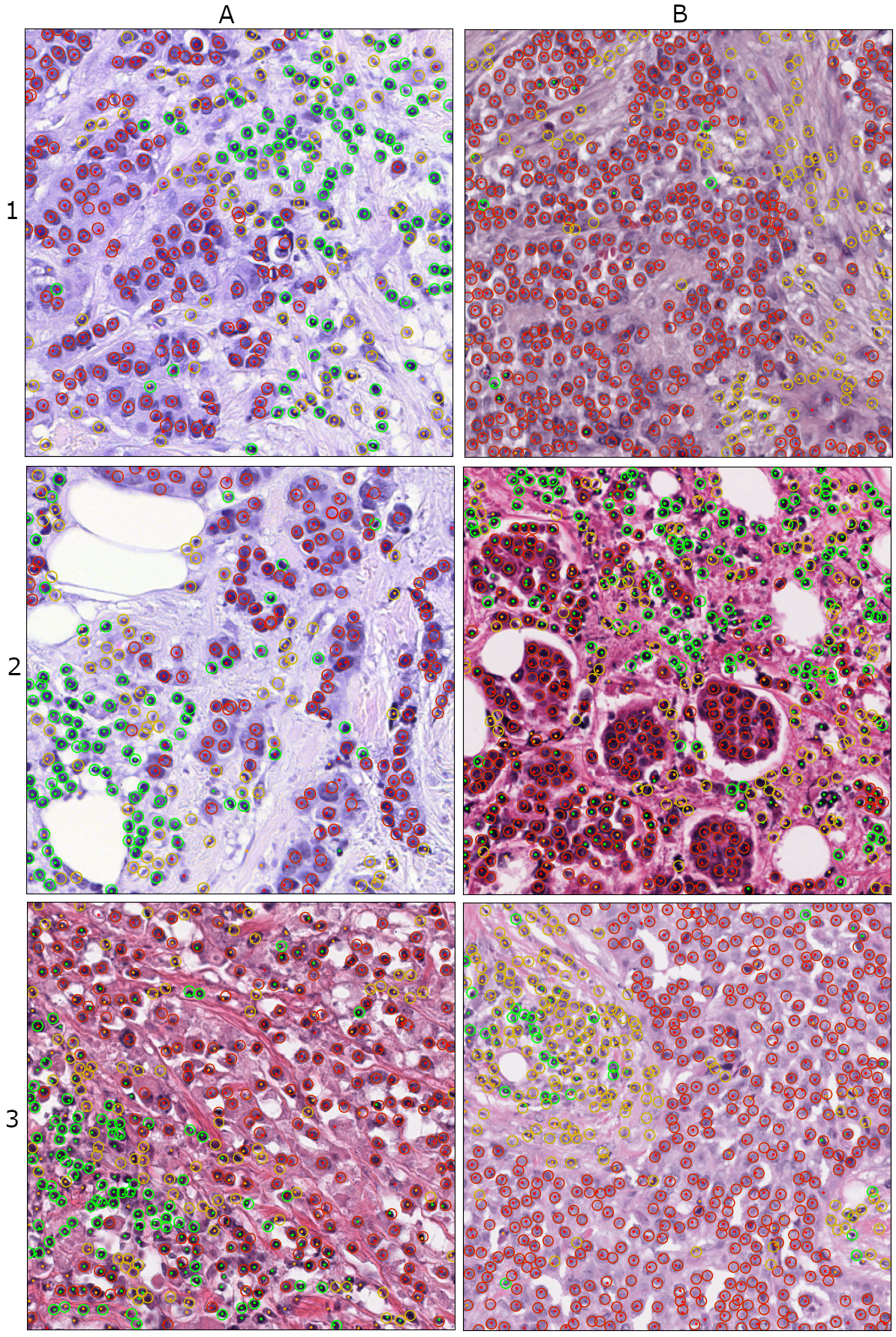}
\caption{Patherea-P2P (ConvNext) qualitative results on BRCA-M2C dataset. Example detections are displayed as filled circles, color-coded for cell type ($\greendot\; lymphocyte$, $\reddot\; epithelial$, $\orangedot\; stromal$). Ground-truth circular region is color coded (border) with the ground-truth cell type. Best viewed in an online version with zoom.}
\label{fig_collage_brca_m2c}
\end{figure}

In Figures~\ref{fig_collage_patherea_2},~\ref{fig_collage_lizard} and~\ref{fig_collage_brca_m2c} we present qualitative results of different approaches and datasets. Figure~\ref{fig_collage_patherea_2} extends Patherea-LNET qualitative results from Figure~\ref{fig_collage_patherea}, by adding the qualitative results from the competing approaches. Figures~\ref{fig_collage_lizard} and~\ref{fig_collage_brca_m2c} show qualitative results of our proposed approach on Lizard~\citep{graham2021lizard} and BRCA-M2C~\citep{abousamra2021multi}. Lizard samples are collected from various public colorectal cancer datasets with various image sizes and resolutions. BRCA-M2C samples are collected and labeled from TCGA-BRCA~\citep{lingle2016cancer}.

\newpage
\bibliographystyle{elsarticle-harv} 
\bibliography{references}

\begin{thebibliography}{65}
\expandafter\ifx\csname natexlab\endcsname\relax\def\natexlab#1{#1}\fi
\providecommand{\url}[1]{\texttt{#1}}
\providecommand{\href}[2]{#2}
\providecommand{\path}[1]{#1}
\providecommand{\DOIprefix}{doi:}
\providecommand{\ArXivprefix}{arXiv:}
\providecommand{\URLprefix}{URL: }
\providecommand{\Pubmedprefix}{pmid:}
\providecommand{\doi}[1]{\href{http://dx.doi.org/#1}{\path{#1}}}
\providecommand{\Pubmed}[1]{\href{pmid:#1}{\path{#1}}}
\providecommand{\bibinfo}[2]{#2}
\ifx\xfnm\relax \def\xfnm[#1]{\unskip,\space#1}\fi
\bibitem[{Abousamra et~al.(2021)Abousamra, Belinsky, Van~Arnam, Allard, Yee,
  Gupta, Kurc, Samaras, Saltz and Chen}]{abousamra2021multi}
\bibinfo{author}{Abousamra, S.}, \bibinfo{author}{Belinsky, D.},
  \bibinfo{author}{Van~Arnam, J.}, \bibinfo{author}{Allard, F.},
  \bibinfo{author}{Yee, E.}, \bibinfo{author}{Gupta, R.},
  \bibinfo{author}{Kurc, T.}, \bibinfo{author}{Samaras, D.},
  \bibinfo{author}{Saltz, J.}, \bibinfo{author}{Chen, C.},
  \bibinfo{year}{2021}.
\newblock \bibinfo{title}{Multi-class cell detection using spatial context
  representation}, in: \bibinfo{booktitle}{Proceedings of the IEEE/CVF
  International Conference on Computer Vision}, pp.
  \bibinfo{pages}{4005--4014}.
\bibitem[{Azizi et~al.(2021)Azizi, Mustafa, Ryan, Beaver, Freyberg, Deaton,
  Loh, Karthikesalingam, Kornblith, Chen et~al.}]{azizi2021big}
\bibinfo{author}{Azizi, S.}, \bibinfo{author}{Mustafa, B.},
  \bibinfo{author}{Ryan, F.}, \bibinfo{author}{Beaver, Z.},
  \bibinfo{author}{Freyberg, J.}, \bibinfo{author}{Deaton, J.},
  \bibinfo{author}{Loh, A.}, \bibinfo{author}{Karthikesalingam, A.},
  \bibinfo{author}{Kornblith, S.}, \bibinfo{author}{Chen, T.}, et~al.,
  \bibinfo{year}{2021}.
\newblock \bibinfo{title}{Big self-supervised models advance medical image
  classification}, in: \bibinfo{booktitle}{Proceedings of the IEEE/CVF
  international conference on computer vision}, pp.
  \bibinfo{pages}{3478--3488}.
\bibitem[{Bankhead et~al.(2017)Bankhead, Loughrey, Fern{\'a}ndez, Dombrowski,
  McArt, Dunne, McQuaid, Gray, Murray, Coleman et~al.}]{bankhead2017qupath}
\bibinfo{author}{Bankhead, P.}, \bibinfo{author}{Loughrey, M.B.},
  \bibinfo{author}{Fern{\'a}ndez, J.A.}, \bibinfo{author}{Dombrowski, Y.},
  \bibinfo{author}{McArt, D.G.}, \bibinfo{author}{Dunne, P.D.},
  \bibinfo{author}{McQuaid, S.}, \bibinfo{author}{Gray, R.T.},
  \bibinfo{author}{Murray, L.J.}, \bibinfo{author}{Coleman, H.G.}, et~al.,
  \bibinfo{year}{2017}.
\newblock \bibinfo{title}{Qupath: Open source software for digital pathology
  image analysis}.
\newblock \bibinfo{journal}{Scientific reports} \bibinfo{volume}{7},
  \bibinfo{pages}{1--7}.
\bibitem[{Bejnordi et~al.(2017)Bejnordi, Veta, Van~Diest, Van~Ginneken,
  Karssemeijer, Litjens, Van Der~Laak, Hermsen, Manson, Balkenhol
  et~al.}]{bejnordi2017diagnostic}
\bibinfo{author}{Bejnordi, B.E.}, \bibinfo{author}{Veta, M.},
  \bibinfo{author}{Van~Diest, P.J.}, \bibinfo{author}{Van~Ginneken, B.},
  \bibinfo{author}{Karssemeijer, N.}, \bibinfo{author}{Litjens, G.},
  \bibinfo{author}{Van Der~Laak, J.A.}, \bibinfo{author}{Hermsen, M.},
  \bibinfo{author}{Manson, Q.F.}, \bibinfo{author}{Balkenhol, M.}, et~al.,
  \bibinfo{year}{2017}.
\newblock \bibinfo{title}{Diagnostic assessment of deep learning algorithms for
  detection of lymph node metastases in women with breast cancer}.
\newblock \bibinfo{journal}{Jama} \bibinfo{volume}{318},
  \bibinfo{pages}{2199--2210}.
\bibitem[{Bera et~al.(2019)Bera, Schalper, Rimm, Velcheti and
  Madabhushi}]{bera2019artificial}
\bibinfo{author}{Bera, K.}, \bibinfo{author}{Schalper, K.A.},
  \bibinfo{author}{Rimm, D.L.}, \bibinfo{author}{Velcheti, V.},
  \bibinfo{author}{Madabhushi, A.}, \bibinfo{year}{2019}.
\newblock \bibinfo{title}{Artificial intelligence in digital pathology—new
  tools for diagnosis and precision oncology}.
\newblock \bibinfo{journal}{Nature reviews Clinical oncology}
  \bibinfo{volume}{16}, \bibinfo{pages}{703--715}.
\bibitem[{Bulten et~al.(2022)Bulten, Kartasalo, Chen, Str{\"o}m, Pinckaers,
  Nagpal, Cai, Steiner, Van~Boven, Vink et~al.}]{bulten2022artificial}
\bibinfo{author}{Bulten, W.}, \bibinfo{author}{Kartasalo, K.},
  \bibinfo{author}{Chen, P.H.C.}, \bibinfo{author}{Str{\"o}m, P.},
  \bibinfo{author}{Pinckaers, H.}, \bibinfo{author}{Nagpal, K.},
  \bibinfo{author}{Cai, Y.}, \bibinfo{author}{Steiner, D.F.},
  \bibinfo{author}{Van~Boven, H.}, \bibinfo{author}{Vink, R.}, et~al.,
  \bibinfo{year}{2022}.
\newblock \bibinfo{title}{Artificial intelligence for diagnosis and gleason
  grading of prostate cancer: the panda challenge}.
\newblock \bibinfo{journal}{Nature medicine} \bibinfo{volume}{28},
  \bibinfo{pages}{154--163}.
\bibitem[{Campanella et~al.(2024)Campanella, Chen, Verma, Zeng, Stock, Croken,
  Veremis, Elmas, Huang, Kwan et~al.}]{campanella2024clinical}
\bibinfo{author}{Campanella, G.}, \bibinfo{author}{Chen, S.},
  \bibinfo{author}{Verma, R.}, \bibinfo{author}{Zeng, J.},
  \bibinfo{author}{Stock, A.}, \bibinfo{author}{Croken, M.},
  \bibinfo{author}{Veremis, B.}, \bibinfo{author}{Elmas, A.},
  \bibinfo{author}{Huang, K.l.}, \bibinfo{author}{Kwan, R.}, et~al.,
  \bibinfo{year}{2024}.
\newblock \bibinfo{title}{A clinical benchmark of public self-supervised
  pathology foundation models}.
\newblock \bibinfo{journal}{arXiv preprint arXiv:2407.06508} .
\bibitem[{Carion et~al.(2020)Carion, Massa, Synnaeve, Usunier, Kirillov and
  Zagoruyko}]{carion2020end}
\bibinfo{author}{Carion, N.}, \bibinfo{author}{Massa, F.},
  \bibinfo{author}{Synnaeve, G.}, \bibinfo{author}{Usunier, N.},
  \bibinfo{author}{Kirillov, A.}, \bibinfo{author}{Zagoruyko, S.},
  \bibinfo{year}{2020}.
\newblock \bibinfo{title}{End-to-end object detection with transformers}, in:
  \bibinfo{booktitle}{European conference on computer vision},
  \bibinfo{organization}{Springer}. pp. \bibinfo{pages}{213--229}.
\bibitem[{Caron et~al.(2021)Caron, Touvron, Misra, J{\'e}gou, Mairal,
  Bojanowski and Joulin}]{caron2021emerging}
\bibinfo{author}{Caron, M.}, \bibinfo{author}{Touvron, H.},
  \bibinfo{author}{Misra, I.}, \bibinfo{author}{J{\'e}gou, H.},
  \bibinfo{author}{Mairal, J.}, \bibinfo{author}{Bojanowski, P.},
  \bibinfo{author}{Joulin, A.}, \bibinfo{year}{2021}.
\newblock \bibinfo{title}{Emerging properties in self-supervised vision
  transformers}, in: \bibinfo{booktitle}{Proceedings of the IEEE/CVF
  international conference on computer vision}, pp.
  \bibinfo{pages}{9650--9660}.
\bibitem[{Chen et~al.(2024)Chen, Ding, Lu, Williamson, Jaume, Song, Chen,
  Zhang, Shao, Shaban et~al.}]{chen2024towards}
\bibinfo{author}{Chen, R.J.}, \bibinfo{author}{Ding, T.}, \bibinfo{author}{Lu,
  M.Y.}, \bibinfo{author}{Williamson, D.F.}, \bibinfo{author}{Jaume, G.},
  \bibinfo{author}{Song, A.H.}, \bibinfo{author}{Chen, B.},
  \bibinfo{author}{Zhang, A.}, \bibinfo{author}{Shao, D.},
  \bibinfo{author}{Shaban, M.}, et~al., \bibinfo{year}{2024}.
\newblock \bibinfo{title}{Towards a general-purpose foundation model for
  computational pathology}.
\newblock \bibinfo{journal}{Nature Medicine} \bibinfo{volume}{30},
  \bibinfo{pages}{850--862}.
\bibitem[{Chen et~al.(2020)Chen, Kornblith, Norouzi and
  Hinton}]{chen2020simple}
\bibinfo{author}{Chen, T.}, \bibinfo{author}{Kornblith, S.},
  \bibinfo{author}{Norouzi, M.}, \bibinfo{author}{Hinton, G.},
  \bibinfo{year}{2020}.
\newblock \bibinfo{title}{A simple framework for contrastive learning of visual
  representations}, in: \bibinfo{booktitle}{International conference on machine
  learning}, \bibinfo{organization}{PMLR}. pp. \bibinfo{pages}{1597--1607}.
\bibitem[{Chen et~al.(2023)Chen, Duan, Wang, He, Lu, Dai and
  Qiao}]{chen2023vision}
\bibinfo{author}{Chen, Z.}, \bibinfo{author}{Duan, Y.}, \bibinfo{author}{Wang,
  W.}, \bibinfo{author}{He, J.}, \bibinfo{author}{Lu, T.},
  \bibinfo{author}{Dai, J.}, \bibinfo{author}{Qiao, Y.}, \bibinfo{year}{2023}.
\newblock \bibinfo{title}{Vision transformer adapter for dense predictions},
  in: \bibinfo{booktitle}{The Eleventh International Conference on Learning
  Representations}.
\bibitem[{Cui and Zhang(2021)}]{cui2021artificial}
\bibinfo{author}{Cui, M.}, \bibinfo{author}{Zhang, D.Y.}, \bibinfo{year}{2021}.
\newblock \bibinfo{title}{Artificial intelligence and computational pathology}.
\newblock \bibinfo{journal}{Laboratory Investigation} \bibinfo{volume}{101},
  \bibinfo{pages}{412--422}.
\bibitem[{Deng et~al.(2009)Deng, Dong, Socher, Li, Li and
  Fei-Fei}]{deng2009imagenet}
\bibinfo{author}{Deng, J.}, \bibinfo{author}{Dong, W.},
  \bibinfo{author}{Socher, R.}, \bibinfo{author}{Li, L.J.},
  \bibinfo{author}{Li, K.}, \bibinfo{author}{Fei-Fei, L.},
  \bibinfo{year}{2009}.
\newblock \bibinfo{title}{Imagenet: A large-scale hierarchical image database},
  in: \bibinfo{booktitle}{2009 IEEE conference on computer vision and pattern
  recognition}, \bibinfo{organization}{Ieee}. pp. \bibinfo{pages}{248--255}.
\bibitem[{Dippel et~al.(2024)Dippel, Feulner, Winterhoff, Schallenberg,
  Dernbach, Kunft, Tietz, Jurmeister, Horst, Ruff et~al.}]{dippel2024rudolfv}
\bibinfo{author}{Dippel, J.}, \bibinfo{author}{Feulner, B.},
  \bibinfo{author}{Winterhoff, T.}, \bibinfo{author}{Schallenberg, S.},
  \bibinfo{author}{Dernbach, G.}, \bibinfo{author}{Kunft, A.},
  \bibinfo{author}{Tietz, S.}, \bibinfo{author}{Jurmeister, P.},
  \bibinfo{author}{Horst, D.}, \bibinfo{author}{Ruff, L.}, et~al.,
  \bibinfo{year}{2024}.
\newblock \bibinfo{title}{{RudolfV: a foundation model by pathologists for
  pathologists}}.
\newblock \bibinfo{journal}{arXiv preprint arXiv:2401.04079} .
\bibitem[{Dosovitskiy et~al.(2021)Dosovitskiy, Beyer, Kolesnikov, Weissenborn,
  Zhai, Unterthiner, Dehghani, Minderer, Heigold, Gelly, Uszkoreit and
  Houlsby}]{dosovitskiy2021an}
\bibinfo{author}{Dosovitskiy, A.}, \bibinfo{author}{Beyer, L.},
  \bibinfo{author}{Kolesnikov, A.}, \bibinfo{author}{Weissenborn, D.},
  \bibinfo{author}{Zhai, X.}, \bibinfo{author}{Unterthiner, T.},
  \bibinfo{author}{Dehghani, M.}, \bibinfo{author}{Minderer, M.},
  \bibinfo{author}{Heigold, G.}, \bibinfo{author}{Gelly, S.},
  \bibinfo{author}{Uszkoreit, J.}, \bibinfo{author}{Houlsby, N.},
  \bibinfo{year}{2021}.
\newblock \bibinfo{title}{An image is worth 16x16 words: Transformers for image
  recognition at scale}, in: \bibinfo{booktitle}{International Conference on
  Learning Representations}.
\bibitem[{Dowsett et~al.(2011)Dowsett, Nielsen, A’Hern, Bartlett, Coombes,
  Cuzick, Ellis, Henry, Hugh, Lively et~al.}]{dowsett2011assessment}
\bibinfo{author}{Dowsett, M.}, \bibinfo{author}{Nielsen, T.O.},
  \bibinfo{author}{A’Hern, R.}, \bibinfo{author}{Bartlett, J.},
  \bibinfo{author}{Coombes, R.C.}, \bibinfo{author}{Cuzick, J.},
  \bibinfo{author}{Ellis, M.}, \bibinfo{author}{Henry, N.L.},
  \bibinfo{author}{Hugh, J.C.}, \bibinfo{author}{Lively, T.}, et~al.,
  \bibinfo{year}{2011}.
\newblock \bibinfo{title}{{Assessment of Ki67 in breast cancer: recommendations
  from the International Ki67 in Breast Cancer working group}}.
\newblock \bibinfo{journal}{Journal of the National cancer Institute}
  \bibinfo{volume}{103}, \bibinfo{pages}{1656--1664}.
\bibitem[{Ericsson et~al.(2021)Ericsson, Gouk and
  Hospedales}]{ericsson2021well}
\bibinfo{author}{Ericsson, L.}, \bibinfo{author}{Gouk, H.},
  \bibinfo{author}{Hospedales, T.M.}, \bibinfo{year}{2021}.
\newblock \bibinfo{title}{How well do self-supervised models transfer?}, in:
  \bibinfo{booktitle}{Proceedings of the IEEE/CVF conference on computer vision
  and pattern recognition}, pp. \bibinfo{pages}{5414--5423}.
\bibitem[{Filiot et~al.(2023)Filiot, Ghermi, Olivier, Jacob, Fidon, Mac~Kain,
  Saillard and Schiratti}]{filiot2023scaling}
\bibinfo{author}{Filiot, A.}, \bibinfo{author}{Ghermi, R.},
  \bibinfo{author}{Olivier, A.}, \bibinfo{author}{Jacob, P.},
  \bibinfo{author}{Fidon, L.}, \bibinfo{author}{Mac~Kain, A.},
  \bibinfo{author}{Saillard, C.}, \bibinfo{author}{Schiratti, J.},
  \bibinfo{year}{2023}.
\newblock \bibinfo{title}{Scaling self-supervised learning for histopathology
  with masked image modeling. medrxiv} .
\bibitem[{Gamper et~al.(2019)Gamper, Alemi~Koohbanani, Benet, Khuram and
  Rajpoot}]{gamper2019pannuke}
\bibinfo{author}{Gamper, J.}, \bibinfo{author}{Alemi~Koohbanani, N.},
  \bibinfo{author}{Benet, K.}, \bibinfo{author}{Khuram, A.},
  \bibinfo{author}{Rajpoot, N.}, \bibinfo{year}{2019}.
\newblock \bibinfo{title}{Pannuke: an open pan-cancer histology dataset for
  nuclei instance segmentation and classification}, in:
  \bibinfo{booktitle}{Digital Pathology: 15th European Congress, ECDP 2019,
  Warwick, UK, April 10--13, 2019, Proceedings 15},
  \bibinfo{organization}{Springer}. pp. \bibinfo{pages}{11--19}.
\bibitem[{Goldblum et~al.(2024)Goldblum, Souri, Ni, Shu, Prabhu, Somepalli,
  Chattopadhyay, Ibrahim, Bardes, Hoffman et~al.}]{goldblum2024battle}
\bibinfo{author}{Goldblum, M.}, \bibinfo{author}{Souri, H.},
  \bibinfo{author}{Ni, R.}, \bibinfo{author}{Shu, M.}, \bibinfo{author}{Prabhu,
  V.}, \bibinfo{author}{Somepalli, G.}, \bibinfo{author}{Chattopadhyay, P.},
  \bibinfo{author}{Ibrahim, M.}, \bibinfo{author}{Bardes, A.},
  \bibinfo{author}{Hoffman, J.}, et~al., \bibinfo{year}{2024}.
\newblock \bibinfo{title}{Battle of the backbones: A large-scale comparison of
  pretrained models across computer vision tasks}.
\newblock \bibinfo{journal}{Advances in Neural Information Processing Systems}
  \bibinfo{volume}{36}.
\bibitem[{Graham et~al.(2021)Graham, Jahanifar, Azam, Nimir, Tsang, Dodd, Hero,
  Sahota, Tank, Benes et~al.}]{graham2021lizard}
\bibinfo{author}{Graham, S.}, \bibinfo{author}{Jahanifar, M.},
  \bibinfo{author}{Azam, A.}, \bibinfo{author}{Nimir, M.},
  \bibinfo{author}{Tsang, Y.W.}, \bibinfo{author}{Dodd, K.},
  \bibinfo{author}{Hero, E.}, \bibinfo{author}{Sahota, H.},
  \bibinfo{author}{Tank, A.}, \bibinfo{author}{Benes, K.}, et~al.,
  \bibinfo{year}{2021}.
\newblock \bibinfo{title}{Lizard: A large-scale dataset for colonic nuclear
  instance segmentation and classification}, in:
  \bibinfo{booktitle}{Proceedings of the IEEE/CVF international conference on
  computer vision}, pp. \bibinfo{pages}{684--693}.
\bibitem[{Graham et~al.(2023)Graham, Vu, Jahanifar, Raza, Minhas, Snead and
  Rajpoot}]{graham2023one}
\bibinfo{author}{Graham, S.}, \bibinfo{author}{Vu, Q.D.},
  \bibinfo{author}{Jahanifar, M.}, \bibinfo{author}{Raza, S.E.A.},
  \bibinfo{author}{Minhas, F.}, \bibinfo{author}{Snead, D.},
  \bibinfo{author}{Rajpoot, N.}, \bibinfo{year}{2023}.
\newblock \bibinfo{title}{One model is all you need: multi-task learning
  enables simultaneous histology image segmentation and classification}.
\newblock \bibinfo{journal}{Medical Image Analysis} \bibinfo{volume}{83},
  \bibinfo{pages}{102685}.
\bibitem[{Graham et~al.(2024)Graham, Vu, Jahanifar, Weigert, Schmidt, Zhang,
  Zhang, Yang, Xiang, Wang et~al.}]{graham2024conic}
\bibinfo{author}{Graham, S.}, \bibinfo{author}{Vu, Q.D.},
  \bibinfo{author}{Jahanifar, M.}, \bibinfo{author}{Weigert, M.},
  \bibinfo{author}{Schmidt, U.}, \bibinfo{author}{Zhang, W.},
  \bibinfo{author}{Zhang, J.}, \bibinfo{author}{Yang, S.},
  \bibinfo{author}{Xiang, J.}, \bibinfo{author}{Wang, X.}, et~al.,
  \bibinfo{year}{2024}.
\newblock \bibinfo{title}{Conic challenge: Pushing the frontiers of nuclear
  detection, segmentation, classification and counting}.
\newblock \bibinfo{journal}{Medical image analysis} \bibinfo{volume}{92},
  \bibinfo{pages}{103047}.
\bibitem[{Graham et~al.(2019)Graham, Vu, Raza, Azam, Tsang, Kwak and
  Rajpoot}]{graham2019hover}
\bibinfo{author}{Graham, S.}, \bibinfo{author}{Vu, Q.D.},
  \bibinfo{author}{Raza, S.E.A.}, \bibinfo{author}{Azam, A.},
  \bibinfo{author}{Tsang, Y.W.}, \bibinfo{author}{Kwak, J.T.},
  \bibinfo{author}{Rajpoot, N.}, \bibinfo{year}{2019}.
\newblock \bibinfo{title}{Hover-net: Simultaneous segmentation and
  classification of nuclei in multi-tissue histology images}.
\newblock \bibinfo{journal}{Medical image analysis} \bibinfo{volume}{58},
  \bibinfo{pages}{101563}.
\bibitem[{Gutman et~al.(2017)Gutman, Khalilia, Lee, Nalisnik, Mullen, Beezley,
  Chittajallu, Manthey and Cooper}]{gutman2017digital}
\bibinfo{author}{Gutman, D.A.}, \bibinfo{author}{Khalilia, M.},
  \bibinfo{author}{Lee, S.}, \bibinfo{author}{Nalisnik, M.},
  \bibinfo{author}{Mullen, Z.}, \bibinfo{author}{Beezley, J.},
  \bibinfo{author}{Chittajallu, D.R.}, \bibinfo{author}{Manthey, D.},
  \bibinfo{author}{Cooper, L.A.}, \bibinfo{year}{2017}.
\newblock \bibinfo{title}{The digital slide archive: a software platform for
  management, integration, and analysis of histology for cancer research}.
\newblock \bibinfo{journal}{Cancer research} \bibinfo{volume}{77},
  \bibinfo{pages}{e75--e78}.
\bibitem[{He et~al.(2022)He, Chen, Xie, Li, Doll{\'a}r and
  Girshick}]{he2022masked}
\bibinfo{author}{He, K.}, \bibinfo{author}{Chen, X.}, \bibinfo{author}{Xie,
  S.}, \bibinfo{author}{Li, Y.}, \bibinfo{author}{Doll{\'a}r, P.},
  \bibinfo{author}{Girshick, R.}, \bibinfo{year}{2022}.
\newblock \bibinfo{title}{Masked autoencoders are scalable vision learners},
  in: \bibinfo{booktitle}{Proceedings of the IEEE/CVF conference on computer
  vision and pattern recognition}, pp. \bibinfo{pages}{16000--16009}.
\bibitem[{He et~al.(2016)He, Zhang, Ren and Sun}]{he2016deep}
\bibinfo{author}{He, K.}, \bibinfo{author}{Zhang, X.}, \bibinfo{author}{Ren,
  S.}, \bibinfo{author}{Sun, J.}, \bibinfo{year}{2016}.
\newblock \bibinfo{title}{Deep residual learning for image recognition}, in:
  \bibinfo{booktitle}{Proceedings of the IEEE conference on computer vision and
  pattern recognition}, pp. \bibinfo{pages}{770--778}.
\bibitem[{H{\"o}rst et~al.(2025)H{\"o}rst, Rempe, Becker, Heine, Keyl and
  Kleesiek}]{cellvit++}
\bibinfo{author}{H{\"o}rst, F.}, \bibinfo{author}{Rempe, M.},
  \bibinfo{author}{Becker, H.}, \bibinfo{author}{Heine, L.},
  \bibinfo{author}{Keyl, J.}, \bibinfo{author}{Kleesiek, J.},
  \bibinfo{year}{2025}.
\newblock \bibinfo{title}{Cellvit++: Energy-efficient and adaptive cell
  segmentation and classification using foundation models}.
\newblock \bibinfo{journal}{arXiv preprint arXiv:2501.05269} .
\bibitem[{H{\"o}rst et~al.(2024)H{\"o}rst, Rempe, Heine, Seibold, Keyl,
  Baldini, Ugurel, Siveke, Gr{\"u}nwald, Egger et~al.}]{cellvit}
\bibinfo{author}{H{\"o}rst, F.}, \bibinfo{author}{Rempe, M.},
  \bibinfo{author}{Heine, L.}, \bibinfo{author}{Seibold, C.},
  \bibinfo{author}{Keyl, J.}, \bibinfo{author}{Baldini, G.},
  \bibinfo{author}{Ugurel, S.}, \bibinfo{author}{Siveke, J.},
  \bibinfo{author}{Gr{\"u}nwald, B.}, \bibinfo{author}{Egger, J.}, et~al.,
  \bibinfo{year}{2024}.
\newblock \bibinfo{title}{Cellvit: Vision transformers for precise cell
  segmentation and classification}.
\newblock \bibinfo{journal}{Medical Image Analysis} \bibinfo{volume}{94},
  \bibinfo{pages}{103143}.
\bibitem[{Huang et~al.(2023a)Huang, Li, Sun, Wan and Li}]{huang2023prompt}
\bibinfo{author}{Huang, J.}, \bibinfo{author}{Li, H.}, \bibinfo{author}{Sun,
  W.}, \bibinfo{author}{Wan, X.}, \bibinfo{author}{Li, G.},
  \bibinfo{year}{2023}a.
\newblock \bibinfo{title}{Prompt-based grouping transformer for nucleus
  detection and classification}, in: \bibinfo{booktitle}{International
  Conference on Medical Image Computing and Computer-Assisted Intervention},
  \bibinfo{organization}{Springer}. pp. \bibinfo{pages}{569--579}.
\bibitem[{Huang et~al.(2023b)Huang, Li, Wan and Li}]{huang2023affine}
\bibinfo{author}{Huang, J.}, \bibinfo{author}{Li, H.}, \bibinfo{author}{Wan,
  X.}, \bibinfo{author}{Li, G.}, \bibinfo{year}{2023}b.
\newblock \bibinfo{title}{Affine-consistent transformer for multi-class cell
  nuclei detection}, in: \bibinfo{booktitle}{Proceedings of the IEEE/CVF
  International Conference on Computer Vision}, pp.
  \bibinfo{pages}{21384--21393}.
\bibitem[{Huang et~al.(2020)Huang, Ding, Song, Wang, Geng, He, Du, Liu, Tian,
  Liang et~al.}]{huang2020bcdata}
\bibinfo{author}{Huang, Z.}, \bibinfo{author}{Ding, Y.}, \bibinfo{author}{Song,
  G.}, \bibinfo{author}{Wang, L.}, \bibinfo{author}{Geng, R.},
  \bibinfo{author}{He, H.}, \bibinfo{author}{Du, S.}, \bibinfo{author}{Liu,
  X.}, \bibinfo{author}{Tian, Y.}, \bibinfo{author}{Liang, Y.}, et~al.,
  \bibinfo{year}{2020}.
\newblock \bibinfo{title}{Bcdata: A large-scale dataset and benchmark for cell
  detection and counting}, in: \bibinfo{booktitle}{Medical Image Computing and
  Computer Assisted Intervention--MICCAI 2020: 23rd International Conference,
  Lima, Peru, October 4--8, 2020, Proceedings, Part V 23},
  \bibinfo{organization}{Springer}. pp. \bibinfo{pages}{289--298}.
\bibitem[{Kainz et~al.(2015)Kainz, Urschler, Schulter, Wohlhart and
  Lepetit}]{kainz2015you}
\bibinfo{author}{Kainz, P.}, \bibinfo{author}{Urschler, M.},
  \bibinfo{author}{Schulter, S.}, \bibinfo{author}{Wohlhart, P.},
  \bibinfo{author}{Lepetit, V.}, \bibinfo{year}{2015}.
\newblock \bibinfo{title}{You should use regression to detect cells}, in:
  \bibinfo{booktitle}{Medical Image Computing and Computer-Assisted
  Intervention--MICCAI 2015: 18th International Conference, Munich, Germany,
  October 5-9, 2015, Proceedings, Part III 18},
  \bibinfo{organization}{Springer}. pp. \bibinfo{pages}{276--283}.
\bibitem[{Kather et~al.(2018)Kather, Halama and Marx}]{kather2018100}
\bibinfo{author}{Kather, J.N.}, \bibinfo{author}{Halama, N.},
  \bibinfo{author}{Marx, A.}, \bibinfo{year}{2018}.
\newblock \bibinfo{title}{100,000 histological images of human colorectal
  cancer and healthy tissue}.
\newblock \bibinfo{journal}{Zenodo10} \bibinfo{volume}{5281}.
\bibitem[{Lee et~al.(2021)Lee, Shaw, Simpson, Uminsky and
  Garratt}]{lee2021differential}
\bibinfo{author}{Lee, S.M.}, \bibinfo{author}{Shaw, A.},
  \bibinfo{author}{Simpson, J.L.}, \bibinfo{author}{Uminsky, D.},
  \bibinfo{author}{Garratt, L.W.}, \bibinfo{year}{2021}.
\newblock \bibinfo{title}{Differential cell counts using center-point networks
  achieves human-level accuracy and efficiency over segmentation}.
\newblock \bibinfo{journal}{Scientific Reports} \bibinfo{volume}{11},
  \bibinfo{pages}{16917}.
\bibitem[{Li et~al.(2018)Li, Zhang and Chen}]{li2018csrnet}
\bibinfo{author}{Li, Y.}, \bibinfo{author}{Zhang, X.}, \bibinfo{author}{Chen,
  D.}, \bibinfo{year}{2018}.
\newblock \bibinfo{title}{Csrnet: Dilated convolutional neural networks for
  understanding the highly congested scenes}, in:
  \bibinfo{booktitle}{Proceedings of the IEEE conference on computer vision and
  pattern recognition}, pp. \bibinfo{pages}{1091--1100}.
\bibitem[{Lin et~al.(2017)Lin, Doll{\'a}r, Girshick, He, Hariharan and
  Belongie}]{lin2017feature}
\bibinfo{author}{Lin, T.Y.}, \bibinfo{author}{Doll{\'a}r, P.},
  \bibinfo{author}{Girshick, R.}, \bibinfo{author}{He, K.},
  \bibinfo{author}{Hariharan, B.}, \bibinfo{author}{Belongie, S.},
  \bibinfo{year}{2017}.
\newblock \bibinfo{title}{Feature pyramid networks for object detection}, in:
  \bibinfo{booktitle}{Proceedings of the IEEE conference on computer vision and
  pattern recognition}, pp. \bibinfo{pages}{2117--2125}.
\bibitem[{Lingle et~al.(2016)Lingle, Erickson, Zuley, Jarosz, Bonaccio,
  Filippini, Net, Levi, Morris, Figler et~al.}]{lingle2016cancer}
\bibinfo{author}{Lingle, W.}, \bibinfo{author}{Erickson, B.},
  \bibinfo{author}{Zuley, M.}, \bibinfo{author}{Jarosz, R.},
  \bibinfo{author}{Bonaccio, E.}, \bibinfo{author}{Filippini, J.},
  \bibinfo{author}{Net, J.}, \bibinfo{author}{Levi, L.},
  \bibinfo{author}{Morris, E.}, \bibinfo{author}{Figler, G.}, et~al.,
  \bibinfo{year}{2016}.
\newblock \bibinfo{title}{The cancer genome atlas breast invasive carcinoma
  collection (tcga-brca)(version 3)[data set] the cancer imaging archive}.
\newblock \bibinfo{journal}{Cancer Imag Arch} .
\bibitem[{Litjens et~al.(2018)Litjens, Bandi, Ehteshami~Bejnordi, Geessink,
  Balkenhol, Bult, Halilovic, Hermsen, Van~de Loo, Vogels
  et~al.}]{litjens20181399}
\bibinfo{author}{Litjens, G.}, \bibinfo{author}{Bandi, P.},
  \bibinfo{author}{Ehteshami~Bejnordi, B.}, \bibinfo{author}{Geessink, O.},
  \bibinfo{author}{Balkenhol, M.}, \bibinfo{author}{Bult, P.},
  \bibinfo{author}{Halilovic, A.}, \bibinfo{author}{Hermsen, M.},
  \bibinfo{author}{Van~de Loo, R.}, \bibinfo{author}{Vogels, R.}, et~al.,
  \bibinfo{year}{2018}.
\newblock \bibinfo{title}{1399 h\&e-stained sentinel lymph node sections of
  breast cancer patients: the camelyon dataset}.
\newblock \bibinfo{journal}{GigaScience} \bibinfo{volume}{7},
  \bibinfo{pages}{giy065}.
\bibitem[{Liu et~al.(2022)Liu, Mao, Wu, Feichtenhofer, Darrell and
  Xie}]{liu2022convnet}
\bibinfo{author}{Liu, Z.}, \bibinfo{author}{Mao, H.}, \bibinfo{author}{Wu,
  C.Y.}, \bibinfo{author}{Feichtenhofer, C.}, \bibinfo{author}{Darrell, T.},
  \bibinfo{author}{Xie, S.}, \bibinfo{year}{2022}.
\newblock \bibinfo{title}{A convnet for the 2020s}, in:
  \bibinfo{booktitle}{Proceedings of the IEEE/CVF conference on computer vision
  and pattern recognition}, pp. \bibinfo{pages}{11976--11986}.
\bibitem[{Loshchilov and Hutter(2019)}]{loshchilov2018decoupled}
\bibinfo{author}{Loshchilov, I.}, \bibinfo{author}{Hutter, F.},
  \bibinfo{year}{2019}.
\newblock \bibinfo{title}{Decoupled weight decay regularization}, in:
  \bibinfo{booktitle}{International Conference on Learning Representations}.
\bibitem[{Mikami et~al.(2013)Mikami, Ueno, Yoshimura, Tsuda, Kurosumi, Masuda,
  Horii, Toi and Sasano}]{mikami2013interobserver}
\bibinfo{author}{Mikami, Y.}, \bibinfo{author}{Ueno, T.},
  \bibinfo{author}{Yoshimura, K.}, \bibinfo{author}{Tsuda, H.},
  \bibinfo{author}{Kurosumi, M.}, \bibinfo{author}{Masuda, S.},
  \bibinfo{author}{Horii, R.}, \bibinfo{author}{Toi, M.},
  \bibinfo{author}{Sasano, H.}, \bibinfo{year}{2013}.
\newblock \bibinfo{title}{{Interobserver concordance of Ki67 labeling index in
  breast cancer: Japan Breast Cancer Research Group Ki67 Ring Study}}.
\newblock \bibinfo{journal}{Cancer science} \bibinfo{volume}{104},
  \bibinfo{pages}{1539--1543}.
\bibitem[{Mukhopadhyay et~al.(2018)Mukhopadhyay, Feldman, Abels, Ashfaq,
  Beltaifa, Cacciabeve, Cathro, Cheng, Cooper, Dickey
  et~al.}]{mukhopadhyay2018whol}
\bibinfo{author}{Mukhopadhyay, S.}, \bibinfo{author}{Feldman, M.D.},
  \bibinfo{author}{Abels, E.}, \bibinfo{author}{Ashfaq, R.},
  \bibinfo{author}{Beltaifa, S.}, \bibinfo{author}{Cacciabeve, N.G.},
  \bibinfo{author}{Cathro, H.P.}, \bibinfo{author}{Cheng, L.},
  \bibinfo{author}{Cooper, K.}, \bibinfo{author}{Dickey, G.E.}, et~al.,
  \bibinfo{year}{2018}.
\newblock \bibinfo{title}{Whole slide imaging versus microscopy for primary
  diagnosis in surgical pathology: a multicenter blinded randomized
  noninferiority study of 1992 cases (pivotal study)}.
\newblock \bibinfo{journal}{The American journal of surgical pathology}
  \bibinfo{volume}{42}, \bibinfo{pages}{39--52}.
\bibitem[{Nechaev et~al.(2024)Nechaev, Pchelnikov and
  Ivanova}]{nechaev2024hibou}
\bibinfo{author}{Nechaev, D.}, \bibinfo{author}{Pchelnikov, A.},
  \bibinfo{author}{Ivanova, E.}, \bibinfo{year}{2024}.
\newblock \bibinfo{title}{Hibou: A family of foundational vision transformers
  for pathology}.
\newblock \bibinfo{journal}{arXiv preprint arXiv:2406.05074} .
\bibitem[{Nielsen et~al.(2021)Nielsen, Leung, Rimm, Dodson, Acs, Badve,
  Denkert, Ellis, Fineberg, Flowers et~al.}]{nielsen2021assessment}
\bibinfo{author}{Nielsen, T.O.}, \bibinfo{author}{Leung, S.C.Y.},
  \bibinfo{author}{Rimm, D.L.}, \bibinfo{author}{Dodson, A.},
  \bibinfo{author}{Acs, B.}, \bibinfo{author}{Badve, S.},
  \bibinfo{author}{Denkert, C.}, \bibinfo{author}{Ellis, M.J.},
  \bibinfo{author}{Fineberg, S.}, \bibinfo{author}{Flowers, M.}, et~al.,
  \bibinfo{year}{2021}.
\newblock \bibinfo{title}{{Assessment of Ki67 in breast cancer: updated
  recommendations from the international Ki67 in breast cancer working group}}.
\newblock \bibinfo{journal}{JNCI: Journal of the National Cancer Institute}
  \bibinfo{volume}{113}, \bibinfo{pages}{808--819}.
\bibitem[{Oquab et~al.(2024)Oquab, Darcet, Moutakanni, Vo, Szafraniec,
  Khalidov, Fernandez, HAZIZA, Massa, El-Nouby, Assran, Ballas, Galuba, Howes,
  Huang, Li, Misra, Rabbat, Sharma, Synnaeve, Xu, Jegou, Mairal, Labatut,
  Joulin and Bojanowski}]{oquab2024dinov}
\bibinfo{author}{Oquab, M.}, \bibinfo{author}{Darcet, T.},
  \bibinfo{author}{Moutakanni, T.}, \bibinfo{author}{Vo, H.V.},
  \bibinfo{author}{Szafraniec, M.}, \bibinfo{author}{Khalidov, V.},
  \bibinfo{author}{Fernandez, P.}, \bibinfo{author}{HAZIZA, D.},
  \bibinfo{author}{Massa, F.}, \bibinfo{author}{El-Nouby, A.},
  \bibinfo{author}{Assran, M.}, \bibinfo{author}{Ballas, N.},
  \bibinfo{author}{Galuba, W.}, \bibinfo{author}{Howes, R.},
  \bibinfo{author}{Huang, P.Y.}, \bibinfo{author}{Li, S.W.},
  \bibinfo{author}{Misra, I.}, \bibinfo{author}{Rabbat, M.},
  \bibinfo{author}{Sharma, V.}, \bibinfo{author}{Synnaeve, G.},
  \bibinfo{author}{Xu, H.}, \bibinfo{author}{Jegou, H.},
  \bibinfo{author}{Mairal, J.}, \bibinfo{author}{Labatut, P.},
  \bibinfo{author}{Joulin, A.}, \bibinfo{author}{Bojanowski, P.},
  \bibinfo{year}{2024}.
\newblock \bibinfo{title}{{DINO}v2: Learning robust visual features without
  supervision}.
\newblock \bibinfo{journal}{Transactions on Machine Learning Research} .
\bibitem[{Pina et~al.(2024)Pina, Dorca and Vilaplana}]{pina2024cell}
\bibinfo{author}{Pina, O.}, \bibinfo{author}{Dorca, E.},
  \bibinfo{author}{Vilaplana, V.}, \bibinfo{year}{2024}.
\newblock \bibinfo{title}{{Cell-DETR: Efficient cell detection and
  classification in WSIs with transformers}}, in: \bibinfo{booktitle}{Medical
  Imaging with Deep Learning}.
\bibitem[{Polley et~al.(2013)Polley, Leung, McShane, Gao, Hugh, Mastropasqua,
  Viale, Zabaglo, Penault-Llorca, Bartlett et~al.}]{polley2013international}
\bibinfo{author}{Polley, M.Y.C.}, \bibinfo{author}{Leung, S.C.},
  \bibinfo{author}{McShane, L.M.}, \bibinfo{author}{Gao, D.},
  \bibinfo{author}{Hugh, J.C.}, \bibinfo{author}{Mastropasqua, M.G.},
  \bibinfo{author}{Viale, G.}, \bibinfo{author}{Zabaglo, L.A.},
  \bibinfo{author}{Penault-Llorca, F.}, \bibinfo{author}{Bartlett, J.M.},
  et~al., \bibinfo{year}{2013}.
\newblock \bibinfo{title}{{An international Ki67 reproducibility study}}.
\newblock \bibinfo{journal}{Journal of the National Cancer Institute}
  \bibinfo{volume}{105}, \bibinfo{pages}{1897--1906}.
\bibitem[{Radford et~al.(2021)Radford, Kim, Hallacy, Ramesh, Goh, Agarwal,
  Sastry, Askell, Mishkin, Clark et~al.}]{radford2021learning}
\bibinfo{author}{Radford, A.}, \bibinfo{author}{Kim, J.W.},
  \bibinfo{author}{Hallacy, C.}, \bibinfo{author}{Ramesh, A.},
  \bibinfo{author}{Goh, G.}, \bibinfo{author}{Agarwal, S.},
  \bibinfo{author}{Sastry, G.}, \bibinfo{author}{Askell, A.},
  \bibinfo{author}{Mishkin, P.}, \bibinfo{author}{Clark, J.}, et~al.,
  \bibinfo{year}{2021}.
\newblock \bibinfo{title}{Learning transferable visual models from natural
  language supervision}, in: \bibinfo{booktitle}{International conference on
  machine learning}, \bibinfo{organization}{PMLR}. pp.
  \bibinfo{pages}{8748--8763}.
\bibitem[{Radford et~al.(2019)Radford, Wu, Child, Luan, Amodei, Sutskever
  et~al.}]{radford2019language}
\bibinfo{author}{Radford, A.}, \bibinfo{author}{Wu, J.},
  \bibinfo{author}{Child, R.}, \bibinfo{author}{Luan, D.},
  \bibinfo{author}{Amodei, D.}, \bibinfo{author}{Sutskever, I.}, et~al.,
  \bibinfo{year}{2019}.
\newblock \bibinfo{title}{Language models are unsupervised multitask learners}.
\newblock \bibinfo{journal}{OpenAI blog} \bibinfo{volume}{1},
  \bibinfo{pages}{9}.
\bibitem[{Reid et~al.(2015)Reid, Bagci, Ohike, Saka, Seven, Dursun, Balci,
  Gucer, Jang, Tajiri et~al.}]{reid2015calculation}
\bibinfo{author}{Reid, M.D.}, \bibinfo{author}{Bagci, P.},
  \bibinfo{author}{Ohike, N.}, \bibinfo{author}{Saka, B.},
  \bibinfo{author}{Seven, I.E.}, \bibinfo{author}{Dursun, N.},
  \bibinfo{author}{Balci, S.}, \bibinfo{author}{Gucer, H.},
  \bibinfo{author}{Jang, K.T.}, \bibinfo{author}{Tajiri, T.}, et~al.,
  \bibinfo{year}{2015}.
\newblock \bibinfo{title}{{Calculation of the Ki67 index in pancreatic
  neuroendocrine tumors: a comparative analysis of four counting
  methodologies}}.
\newblock \bibinfo{journal}{Modern Pathology} \bibinfo{volume}{28},
  \bibinfo{pages}{686--694}.
\bibitem[{Ryu et~al.(2023)Ryu, Puche, Shin, Park, Brattoli, Lee, Jung, Cho,
  Paeng, Ock et~al.}]{ryu2023ocelot}
\bibinfo{author}{Ryu, J.}, \bibinfo{author}{Puche, A.V.},
  \bibinfo{author}{Shin, J.}, \bibinfo{author}{Park, S.},
  \bibinfo{author}{Brattoli, B.}, \bibinfo{author}{Lee, J.},
  \bibinfo{author}{Jung, W.}, \bibinfo{author}{Cho, S.I.},
  \bibinfo{author}{Paeng, K.}, \bibinfo{author}{Ock, C.Y.}, et~al.,
  \bibinfo{year}{2023}.
\newblock \bibinfo{title}{Ocelot: overlapped cell on tissue dataset for
  histopathology}, in: \bibinfo{booktitle}{Proceedings of the IEEE/CVF
  Conference on Computer Vision and Pattern Recognition}, pp.
  \bibinfo{pages}{23902--23912}.
\bibitem[{Sirinukunwattana et~al.(2017)Sirinukunwattana, Pluim, Chen, Qi, Heng,
  Guo, Wang, Matuszewski, Bruni, Sanchez et~al.}]{sirinukunwattana2017gland}
\bibinfo{author}{Sirinukunwattana, K.}, \bibinfo{author}{Pluim, J.P.},
  \bibinfo{author}{Chen, H.}, \bibinfo{author}{Qi, X.}, \bibinfo{author}{Heng,
  P.A.}, \bibinfo{author}{Guo, Y.B.}, \bibinfo{author}{Wang, L.Y.},
  \bibinfo{author}{Matuszewski, B.J.}, \bibinfo{author}{Bruni, E.},
  \bibinfo{author}{Sanchez, U.}, et~al., \bibinfo{year}{2017}.
\newblock \bibinfo{title}{Gland segmentation in colon histology images: The
  glas challenge contest}.
\newblock \bibinfo{journal}{Medical image analysis} \bibinfo{volume}{35},
  \bibinfo{pages}{489--502}.
\bibitem[{Sirinukunwattana et~al.(2016)Sirinukunwattana, Raza, Tsang, Snead,
  Cree and Rajpoot}]{sirinukunwattana2016locality}
\bibinfo{author}{Sirinukunwattana, K.}, \bibinfo{author}{Raza, S.E.A.},
  \bibinfo{author}{Tsang, Y.W.}, \bibinfo{author}{Snead, D.R.},
  \bibinfo{author}{Cree, I.A.}, \bibinfo{author}{Rajpoot, N.M.},
  \bibinfo{year}{2016}.
\newblock \bibinfo{title}{Locality sensitive deep learning for detection and
  classification of nuclei in routine colon cancer histology images}.
\newblock \bibinfo{journal}{IEEE transactions on medical imaging}
  \bibinfo{volume}{35}, \bibinfo{pages}{1196--1206}.
\bibitem[{Song et~al.(2021)Song, Wang, Jiang, Wang, Tai, Wang, Li, Huang and
  Wu}]{song2021rethinking}
\bibinfo{author}{Song, Q.}, \bibinfo{author}{Wang, C.}, \bibinfo{author}{Jiang,
  Z.}, \bibinfo{author}{Wang, Y.}, \bibinfo{author}{Tai, Y.},
  \bibinfo{author}{Wang, C.}, \bibinfo{author}{Li, J.}, \bibinfo{author}{Huang,
  F.}, \bibinfo{author}{Wu, Y.}, \bibinfo{year}{2021}.
\newblock \bibinfo{title}{Rethinking counting and localization in crowds: A
  purely point-based framework}, in: \bibinfo{booktitle}{Proceedings of the
  IEEE/CVF International Conference on Computer Vision}, pp.
  \bibinfo{pages}{3365--3374}.
\bibitem[{Tian et~al.(2023)Tian, Jiang, qishuai diao, Lin, Wang and
  Yuan}]{tian2023designing}
\bibinfo{author}{Tian, K.}, \bibinfo{author}{Jiang, Y.},
  \bibinfo{author}{qishuai diao}, \bibinfo{author}{Lin, C.},
  \bibinfo{author}{Wang, L.}, \bibinfo{author}{Yuan, Z.}, \bibinfo{year}{2023}.
\newblock \bibinfo{title}{Designing {BERT} for convolutional networks: Sparse
  and hierarchical masked modeling}, in: \bibinfo{booktitle}{The Eleventh
  International Conference on Learning Representations}.
\bibitem[{Vorontsov et~al.(2023)Vorontsov, Bozkurt, Casson, Shaikovski,
  Zelechowski, Liu, Mathieu, van Eck, Lee, Viret et~al.}]{vorontsov2023virchow}
\bibinfo{author}{Vorontsov, E.}, \bibinfo{author}{Bozkurt, A.},
  \bibinfo{author}{Casson, A.}, \bibinfo{author}{Shaikovski, G.},
  \bibinfo{author}{Zelechowski, M.}, \bibinfo{author}{Liu, S.},
  \bibinfo{author}{Mathieu, P.}, \bibinfo{author}{van Eck, A.},
  \bibinfo{author}{Lee, D.}, \bibinfo{author}{Viret, J.}, et~al.,
  \bibinfo{year}{2023}.
\newblock \bibinfo{title}{Virchow: a million-slide digital pathology foundation
  model}.
\newblock \bibinfo{journal}{arXiv preprint arXiv:2309.07778} .
\bibitem[{Wang et~al.(2020)Wang, Gao, Lin and Li}]{wang2020nwpu}
\bibinfo{author}{Wang, Q.}, \bibinfo{author}{Gao, J.}, \bibinfo{author}{Lin,
  W.}, \bibinfo{author}{Li, X.}, \bibinfo{year}{2020}.
\newblock \bibinfo{title}{Nwpu-crowd: A large-scale benchmark for crowd
  counting and localization}.
\newblock \bibinfo{journal}{IEEE transactions on pattern analysis and machine
  intelligence} \bibinfo{volume}{43}, \bibinfo{pages}{2141--2149}.
\bibitem[{Xie et~al.(2015)Xie, Xing, Kong, Su and Yang}]{xie2015beyond}
\bibinfo{author}{Xie, Y.}, \bibinfo{author}{Xing, F.}, \bibinfo{author}{Kong,
  X.}, \bibinfo{author}{Su, H.}, \bibinfo{author}{Yang, L.},
  \bibinfo{year}{2015}.
\newblock \bibinfo{title}{Beyond classification: structured regression for
  robust cell detection using convolutional neural network}, in:
  \bibinfo{booktitle}{Medical Image Computing and Computer-Assisted
  Intervention--MICCAI 2015: 18th International Conference, Munich, Germany,
  October 5-9, 2015, Proceedings, Part III 18},
  \bibinfo{organization}{Springer}. pp. \bibinfo{pages}{358--365}.
\bibitem[{Xie et~al.(2018)Xie, Xing, Shi, Kong, Su and Yang}]{xie2018efficient}
\bibinfo{author}{Xie, Y.}, \bibinfo{author}{Xing, F.}, \bibinfo{author}{Shi,
  X.}, \bibinfo{author}{Kong, X.}, \bibinfo{author}{Su, H.},
  \bibinfo{author}{Yang, L.}, \bibinfo{year}{2018}.
\newblock \bibinfo{title}{Efficient and robust cell detection: A structured
  regression approach}.
\newblock \bibinfo{journal}{Medical image analysis} \bibinfo{volume}{44},
  \bibinfo{pages}{245--254}.
\bibitem[{Xie et~al.(2022)Xie, Zhang, Cao, Lin, Bao, Yao, Dai and
  Hu}]{xie2022simmim}
\bibinfo{author}{Xie, Z.}, \bibinfo{author}{Zhang, Z.}, \bibinfo{author}{Cao,
  Y.}, \bibinfo{author}{Lin, Y.}, \bibinfo{author}{Bao, J.},
  \bibinfo{author}{Yao, Z.}, \bibinfo{author}{Dai, Q.}, \bibinfo{author}{Hu,
  H.}, \bibinfo{year}{2022}.
\newblock \bibinfo{title}{Simmim: A simple framework for masked image
  modeling}, in: \bibinfo{booktitle}{Proceedings of the IEEE/CVF conference on
  computer vision and pattern recognition}, pp. \bibinfo{pages}{9653--9663}.
\bibitem[{Xing and Yang(2016)}]{xing2016robust}
\bibinfo{author}{Xing, F.}, \bibinfo{author}{Yang, L.}, \bibinfo{year}{2016}.
\newblock \bibinfo{title}{Robust nucleus/cell detection and segmentation in
  digital pathology and microscopy images: a comprehensive review}.
\newblock \bibinfo{journal}{IEEE reviews in biomedical engineering}
  \bibinfo{volume}{9}, \bibinfo{pages}{234--263}.
\bibitem[{Xu et~al.(2024)Xu, Usuyama, Bagga, Zhang, Rao, Naumann, Wong, Gero,
  Gonz{\'a}lez, Gu et~al.}]{xu2024whole}
\bibinfo{author}{Xu, H.}, \bibinfo{author}{Usuyama, N.},
  \bibinfo{author}{Bagga, J.}, \bibinfo{author}{Zhang, S.},
  \bibinfo{author}{Rao, R.}, \bibinfo{author}{Naumann, T.},
  \bibinfo{author}{Wong, C.}, \bibinfo{author}{Gero, Z.},
  \bibinfo{author}{Gonz{\'a}lez, J.}, \bibinfo{author}{Gu, Y.}, et~al.,
  \bibinfo{year}{2024}.
\newblock \bibinfo{title}{A whole-slide foundation model for digital pathology
  from real-world data}.
\newblock \bibinfo{journal}{Nature} , \bibinfo{pages}{1--8}.
\bibitem[{Yosinski et~al.(2014)Yosinski, Clune, Bengio and
  Lipson}]{yosinski2014transferable}
\bibinfo{author}{Yosinski, J.}, \bibinfo{author}{Clune, J.},
  \bibinfo{author}{Bengio, Y.}, \bibinfo{author}{Lipson, H.},
  \bibinfo{year}{2014}.
\newblock \bibinfo{title}{How transferable are features in deep neural
  networks?}
\newblock \bibinfo{journal}{Advances in neural information processing systems}
  \bibinfo{volume}{27}.

\end{thebibliography}
\end{document}